\documentclass[10pt,journal,twoside,onecolumn,romanappendices]{IEEEtran}
\usepackage{amsmath,graphicx}
\usepackage{cite}
\usepackage{mathrsfs}
\usepackage{algorithm}
\usepackage{algorithmic}
\usepackage{setspace}
\usepackage{epstopdf}
\usepackage{multirow}
\usepackage{amsfonts}

\usepackage{color}

\usepackage{lineno}

\begin{document}

%\title{ {Learning Deep Priors and Adjusting Its Contribution for Hyperspectral Image Super-Resolution}}
\title{{Hyperspectral Image Super-Resolution via Deep Prior Regularization with Parameter Estimation}}
%\title{Learning and Adjusting Deep Prior for Hyperspectral Image Super-Resolution}
%\title{When Deep Learning meets Optimization Techniques: Super-Resolution of Hyperspectral Image}
%

\author{Xiuheng~Wang,~\IEEEmembership{Student Member,~IEEE,}
        Jie~Chen,~\IEEEmembership{Senior Member,~IEEE,}
        Qi~Wei,~\IEEEmembership{Member,~IEEE,}
        and C\'edric~Richard,~\IEEEmembership{Senior Member,~IEEE,}% <-this % stops a space
\thanks{%Manuscript received September 15, 2018; revised December 23, 2018 and February 19, 2019; accepted March 7, 2019.
X. Wang and J. Chen are with with Centre of Intelligent Acoustics and Immersive Communications at School of Marine Science and Technology, Northwestern Polytechnical University, Xi'an, and also with China Key Laboratory of Ocean Acoustics and Sensing, Ministry of Industry and Information Technology, Xi’an 710072, China (e-mail: xiuheng.wang@mail.nwpu.edu.cn; dr.jie.chen@ieee.org).

Q. Wei is with University of Toulouse, IRIT/INP-ENSEEIHT, 2 rue Camichel, BP 7122, 31071 Toulouse cedex 7, France. (e-mail: bjweiqi@gmail.com).

C. Richard is with Universit\'{e} C\^{o}te d'Azur, OCA, CNRS, 75016 Paris, France (cedric.richard@unice.fr).}}

\maketitle

\begin{abstract}
Hyperspectral image (HSI) super-resolution is commonly used to overcome the hardware limitations of existing hyperspectral imaging systems on spatial resolution. It fuses a low-resolution (LR) HSI and a high-resolution (HR) conventional image of the same scene to obtain an HR HSI. In this work, we propose a method that integrates a physical model and deep prior information. Specifically, a novel, yet effective two-stream fusion network is designed to serve as a  {regularizer} for the fusion problem. This fusion problem is formulated as an optimization problem whose solution can be obtained by solving a Sylvester equation. Furthermore, the regularization parameter is simultaneously estimated to automatically adjust contribution of the physical model and  {the} learned prior to reconstruct the final HR HSI. Experimental results on  {both simulated and real data} demonstrate the superiority of the proposed method over other state-of-the-art methods on both quantitative and qualitative comparisons.

%This technique involves an effective image priors structure required to regularize an ill-posed problem with a proper integration into the degradation model. In this paper, we present a novel, yet effective two-stream fusion nerk (TSFN) in order to directly learn deep priors of the latent HR HSI by means of fusing two input images whose output serves as a regularization term of the fusion problem. Then, we optimize the fusion problem by solving a Sylvester equation, and simultaneously estimate the regularization parameter to automatically to reconstruct the final HR HSI. Experimental results on two public datasets demonstrate the superiority of the proposed method over several state-of-the-art methods on both quantitative and qualitative comparisons. 
\end{abstract}


\section{Introduction}
\IEEEPARstart{H}{yperspectral} imaging simultaneously captures images of the same scene at different wavelengths. Rich spectral characteristics provided by hyperspectral images are important in remote sensing~\cite{bioucas2013hyperspectral}.  {HSIs have been demonstrated to improve the performance of tasks including tracking~\cite{van2010tracking}, classification~\cite{fauvel2012advances, xie2020multiscale}, segmentation~\cite{tarabalka2010segmentation} and clustering~\cite{lei2020deep}.} HSIs contain richer spectral information of real scenes compared with conventional images such as color or gray-scale images. 
However, the high spectral resolution of hyperspectral images  {needs} to make a compromise with spatial resolution to ensure an acceptable signal-to-noise ratio (SNR)~\cite{lanaras2015hyperspectral}. This in turn restricts applications of HSIs. To cope with this issue, a critical task is thus to improve the spatial resolution of HSIs.  {Besides, conventional approaches such as push-broom spectral imaging sensors are often time-consuming and limit applications of HSIs. Recently, researchers designed dual-camera compressive hyperspectral imagers that can capture full HSIs with a single signal exposure, and acquire hyperspectral videos with successive exposures. These techniques reconstruct a full hyperspectral cube relying on the compressive sensing theory and computational imaging algorithms; detailed discussions can be found in~\cite{wang2016simultaneous, zhang2018fast, xu2020hyperspectral}.}

Due to  various physical limitations at hardware level, it is often challenging to develop hyperspectral cameras that simultaneously achieve high spatial and high spectral resolutions. On the other hand, conventional cameras capture RGB (red, green and blue) or panchromatic images with higher spatial resolution but lower spectral resolution by integrating the scene response over broad spectral bands~\cite{lanaras2015hyperspectral}. Consequently, researchers have proposed computational methods to restore a high-resolution (HR) HSI by integrating a low-resolution (LR) HSI and an HR conventional image~\cite{alparone2007comparison, yokoya2017hyperspectral}. This procedure is often referred to as \textit{HSI super-resolution} or \textit{HSI fusion}. 
A class of fusion approaches is based on component substitution such as the intensity, hue and saturation (IHS) technique~\cite{carper1990use}, principal component analysis (PCA)~\cite{kwarteng1989extracting} and wavelet transform~\cite{shettigara1992generalized, nunez1999multiresolution}. These  methods are fast but tend to introduce spectral distortion when estimating the scene response of the latent HR HSIs from panchromatic images. 

It has been popular to formulate the fusion problem as an image restoration problem~\cite{zhang2018exploiting,yokoya2011coupled,lanaras2015hyperspectral,wei2016multiband,huang2013spatial,akhtar2014sparse,wei2015hyperspectral,dong2016hyperspectral, dian2017hyperspectral, li2018fusing, dian2019learning,bu2020hyperspectral}. In such approaches, following a physical degradation model, the input LR HSI and HR conventional images are seen as the spatially and spectrally degraded observations (i.e. linear down-sampled versions) of the latent HR HSI respectively. 
Note that HSI super-resolution is a highly ill-posed problem owing to the large scaling factors in both spatial and spectral domain.  {Thus}, it is important to incorporate prior information to constrain the solution space.  {Depending on the structures of prior information used in the optimization problem, existing techniques can be roughly divided into three categories: spectral unmixing based approaches~\cite{yokoya2011coupled,lanaras2015hyperspectral,wei2016multiband}, sparse representation based approaches~\cite{huang2013spatial,akhtar2014sparse,wei2015hyperspectral,dong2016hyperspectral} and tensor factorization based approaches~\cite{dian2017hyperspectral, li2018fusing, dian2019learning,bu2020hyperspectral}. A brief review on those methods is given in Section~\ref{section:related_work}.} However, these predefined priors-based methods with handcrafted regularizers have some inherent disadvantages e.g., a complex regularizer may introduce extra difficulties in solving the related optimization problem. %A critical point is \textbf{whether we can automate the construction of a regularizer.} 

Recently, inspired by the successful application of deep learning to many computer vision tasks, especially single image super-resolution (SISR)~\cite{yang2019deep}, convolutional neural networks (CNNs) have been introduced to address the {HSI super-resolution problem~\cite{palsson2017multispectral, yang2017pannet, scarpa2018target, li2019joint, deng2020deep,shi2018deep}}. Compared with optimization methods based on predefined priors, these deep learning methods require fewer assumptions on the prior knowledge of the latent HR HSI, and can  directly learn the relevant information from training data in an end-to-end way. However, these learning based methods ignore the blurring and down-sampling operators as well as the spectral response function in the degradation model, though this model has a clear physical interpretation that relates the LR HSI and the HR conventional image to the HR HSI.  {To tackle this issue, recent deep learning methods~\cite{xie2019multispectral, zhang2020unsupervised} successfully improved the quality of generated HR HSIs by learning the blurring matrix and the spectral response function. The work in~\cite{xie2019multispectral} formulated a fusion model considering both spatial and spectral degradation. A deep network called MHF-net was accordingly proposed to solve this problem in an iterative manner. In~\cite{zhang2020unsupervised}, a two-stage network based on unsupervised adaption learning (UAL) was proposed to learn the deep priors and estimate the unknown spatial degradation.}

%Based on the degradation model, predefined priors-based optimization methods need to handcraft powerful regulariziers. On the other hand, deep learning methods avoid the assumption of the prior knowledge but neglect the operators in the degradation model. 
In order to leverage the advantages of both optimization and deep learning methods, several deep priors-based approaches (i.e., learning the priors via deep CNNs) have been recently proposed ~\cite{dian2018deep, xie2019hyperspectral}. 
In these methods, deep neural networks were designed to exploit both spatial and spectral characteristics of the latent HR HSI, and their outputs then served as regularizers. This allowed these approaches to achieve enhanced performance. However, they learned deep priors from the images produced by solving a Sylvester equation~\cite{bartels1972solution} rather than two observed images. Thus, the efficacy of deep priors relied on the accuracy of the produced images in their HSI super-resolution schemes. %\textbf{Is there a more suitable network architecture to learn deep priors directly from the two observed images?}

Although numerous works have explored the prior structures of the latent HR HSI, few of them have investigated the way of balancing the contribution of the physical model and prior information in their frameworks. That is however a key point for the effective implementation of such methods. % \textbf{How to balance contribution of the extracted image priors?}
%Since properly adjusting the regularization parameter is one useful way to balance the modeling error and regularization strength. 
 {Some classic regularization parameter estimation methods are reviewed in Section~\ref{section:related_work}.}

Here, we aim at performing hyperspectral super-resolution by combining the physical model-based optimization and deep learning for constructing data-driven priors. Before proceeding with the algorithm, we raise the following three critical points to guide the problem formulation and solving:
\begin{itemize}
     \item How to design a proper deep neural network capable of learning prior information from multi-source data?
     \item How to formulate the fusion problem in a way that can leverage both the physical model and end-to-end learned prior?
     \item How to balance the contribution of the physical model and prior learned from data?
\end{itemize}
% to avoid defining the prior knowledge of the latent HR HSI, and meanwhile consider the degradation model. 
In this work, a new deep priors-based method is proposed to practically
address the three issues pointed above.  {We summarize our contributions as follows:}
\begin{itemize}
\item To avoid handcrafting regularizers as in existing predefined priors-based methods, we train a new effective two-stream fusion network (TSFN) to directly learn deep priors through the fusion of two input observed images rather than produced images. 
\item In order to integrate deep priors into the degradation model, the network output, which corresponds to the extracted deep priors, is plugged into the objective function of the fusion problem as a regularization term.  {The squared Euclidean distance is finally selected as the regularizer after many attempts with various regularizers.}
\item To adjust the contribution of the deep priors {in the hyperspectral super-resolution problem}, we  {adopt the minimum distance criterion (MDC)~\cite{song2016regularization} to} the response curve of the bi-objective optimization problem for automatic selection of the regularization parameter {with a golden-section search strategy}. To our best knowledge, this is the first work designed to balance deep learning approach and optimization approach in this mathematical way. 
\item Our experimental results demonstrate the effectiveness of the proposed HSI super-resolution strategy,  offering an improvement over the results of the state-of-the-art methods.
\end{itemize}
The rest of this paper is organized as follows.  {Related literatures are reviewed in Section \uppercase\expandafter{\romannumeral2}. We formulate the HSI super-resolution problem in Section \uppercase\expandafter{\romannumeral3}. In Section \uppercase\expandafter{\romannumeral4}, the proposed method for HSI super-resolution is explained in detail. Section \uppercase\expandafter{\romannumeral5} presents and discusses experimental results on four public datasets and ablation studies. %Ablation studies are introduced in Section \uppercase\expandafter{\romannumeral6}. 
Finally, conclusions and future works are given in Section \uppercase\expandafter{\romannumeral6}.}

\section{ {Related Work}}
\label{section:related_work}
 {In this section, we will briefly review three categories of existing methods related to this work.}
\subsection{ {Spectral Unmixing Based Approaches}}
 {Leveraging} priors for spectral unmixing with some constraints (e.g., non-negativity, sum-to-one)  {has been experimentally demonstrated to be beneficial} for HSI super-resolution. In~\cite{yokoya2011coupled}, the coupled  {non-negative} matrix factorization (CNMF) method was proposed to alternatively  {unmix} an LR HSI and an HR conventional image to estimate an HR HSI. In~\cite{lanaras2015hyperspectral}, with a similar framework, by jointly unmixing two input images, the initial optimization problem was first decoupled into two constrained least-square problems and then solved. The method proposed in~\cite{wei2016multiband} reconstructed the latent HR HSI with respect to its endmembers and their abundances by using the alternating direction method of multipliers (ADMM) technique~\cite{boyd2011distributed}. 
\subsection{ {Sparse Representation Based Approaches}}
Sparse representation is another promising technique for fusing hyperspectral and conventional images, aiming to sparsely encode the latent HR HSI with an appropriate spectral dictionary learned from the input HSI. To this end, in~\cite{huang2013spatial} the spectral dictionary was learned with K-SVD and sparse matrix factorization was used to fuse two input images. Considering the similarity of neighboring pixels in the latent HSI, the method proposed in~\cite{akhtar2014sparse} enforced group sparsity as well as non-negativity among pixels within small patches. The work~\cite{wei2015hyperspectral} fused the hyperspectral and conventional images in a variational approach and designed a sparse regularizer by decomposing the scene on a spectral dictionary. The method of~\cite{dong2016hyperspectral} developed a non-negative sparse coding algorithm that exploited not only the sparsity of each pixel but also the non-local spatial similarity of the latent HR HSI, leading to enhanced fusion performance. 

\subsection{ {Tensor Factorization Based Approaches}}
 {Another important approach in HSI super-resolution is based on tensor factorization, which has been commonly used to simultaneously capture correlations among different modes. The method in~\cite{dian2017hyperspectral} employed non-local sparse tensor factorization to decompose each cube of the HSI as a sparse core tensor and dictionaries of three modes. To exploit the non-local spatial self-similarities, similar cubes were grouped together and assumed to share the same dictionaries. The work~\cite{li2018fusing} formulated the fusion problem as the coupled sparse Tucker
decomposition, which alternatively updated the dictionaries of three modes and sparse core tensor.  In~\cite{dian2019learning}, a low tensor-train rank (LTTR) based method was proposed to learn correlations among the spatial, spectral and non-local modes of the non-local similar HR HSI cubes to regularize the super-resolution problem. More recently, a graph Laplacian-guided coupled tensor decomposition model was proposed in~\cite{bu2020hyperspectral}. It adopted the coupled
Tucker decomposition to exploit jointly and seamlessly
the intrinsic spatial-spectral information of hyperspectral and conventional images.}

\subsection{ {Regularization Parameter Estimation approaches}}
A classical method for estimating  regularization parameters is the generalized cross-validation (GCV)~\cite{golub1979generalized}, which has been applied for image restoration in~\cite{galatsanos1992methods}. In~\cite{hansen1992analysis}, the regularization parameter was chosen by finding a point near the ``corner" of the L-curve, which was formed by ploting the data fidelity term and the regularizer in log-log scale. However, the use of the log-log scale led to a convexity loss of the L-curve~\cite{kaufman1997regularization}. Recently, basis pursuit has been formulated as a constrained least-square problem, and the corresponding $\textit{Pareto front}$~\cite{deb2001multi} has been proved convex and continuously differentiable over all points of interest in~\cite{van2009probing}. Inspired by this method, the work~\cite{song2016regularization} proposed the maximum curvature criterion (MCC) and the MDC on the response surface (i.e., the linear plot of the data fidelity and regularization costs) and confirmed their good performance in estimating  the regularization parameter for non-negative HSI deconvolution.

\begin{figure*}[]
	\centering
	\includegraphics[scale=1]{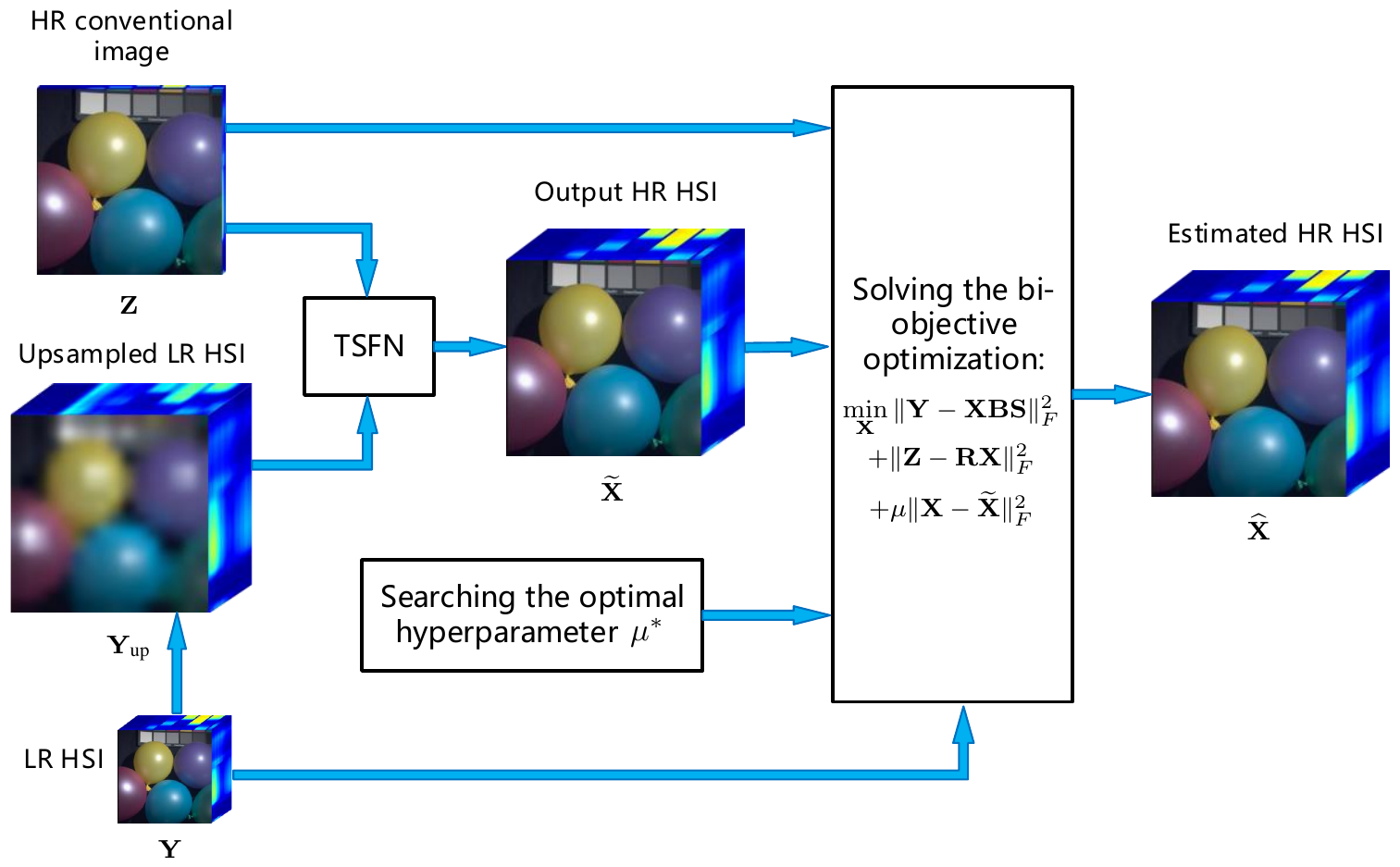}
	\vspace{0mm}
	\caption{Overall scheme of our proposed HSI super-resolution method. HR and LR represent high-resolution and low-resolution in spatial domains, respectively.}
	\label{fig:scheme}
	\vspace{0mm}
\end{figure*}
\section{Problem Formulation}
Consider to reconstruct a high-resolution HSI $X_\text{3D}\in\mathbb{R}^{B\times L \times W}$ based on a low-resolution HSI $Y_\text{3D}\in\mathbb{R}^{B\times l \times w}$ and a high-resolution conventional image $Z_\text{3D}\in\mathbb{R}^{b\times L \times W}$ over the same scenario, where $B$ and $b$ are numbers of spectral bands of the hyperspectral and conventional images ($B > b$); $(L, W)$ and $(l, w)$ are the  {height and width} of the HR and LR images respectively ($L > l, W > w$).
For ease of mathematical formulation, $X_\text{3D}, Y_\text{3D}$ and $Z_\text{3D}$ are transformed in matrix forms $\mathbf{X}\in\mathbb{R}^{B\times N}, \mathbf{Y}\in\mathbb{R}^{B\times n}$ and $\mathbf{Z}\in\mathbb{R}^{b\times N}$, respectively, where $N = L \times W, n = l \times w$ are numbers of pixels in each band of the HR and LR images. According to the linear degradation model, $\mathbf{Y}$ can be viewed as a spatially down-sampled vision of $\mathbf{X}$, while $\mathbf{Z}$ is a down-sampled observation of $\mathbf{X}$ in spectral domain, so that:
\begin{equation}\label{eq:degradation}
	\mathbf{Y} = \mathbf{XBS},\quad
	\mathbf{Z} = \mathbf{RX},
\end{equation}
where $\mathbf{B}\in\mathbb{R}^{N\times N}$ represents the blurring matrix, $\mathbf{S}\in\mathbb{R}^{N\times n}$ is a uniform down-sampling operator with scaling factor $s=N/n$ and $\mathbf{R}\in\mathbb{R}^{b\times B}$ denotes the spectral response function (SRF) of the conventional camera sensor, which is often known or can be estimated $\textit{a priori}$. Moreover, the blurring matrix $\mathbf{B}$ is assumed to be a known block circulant matrix with circulant blocks~\cite{wei2015hyperspectral}. In this widely admitted assumption, $\mathbf{B}$ can be decomposed as
$\mathbf{B} = \mathbf{FDF}^{H}$. Here, $\mathbf{F}\in\mathbb{R}^{N\times N}$ is the discrete Fourier transform (DFT) matrix ($\mathbf{FF}^{H}=\mathbf{I}_{N}$, and $\mathbf{I}_{N}$ is an $N\times N$ identity matrix.) and $\mathbf{D}\in\mathbb{R}^{N\times N}$ is a
diagonal matrix containing the eigenvalues of $\mathbf{B}$. 
 
Based on the degradation model in~\eqref{eq:degradation}, we can estimate $\mathbf{X}$ by introducing proper priors and seeking the minimum of the following bi-objective function:
\begin{equation}\label{eq:object}
\begin{aligned}
	\widehat{\mathbf{X}} & = \arg\mathop{\min}_{{\bf X}}\, \mathcal{J}(\mathbf{X}) \\
	&=  \arg\mathop{\min}_{{\bf X}}\,  \mathcal{J}_1 (\mathbf{X}) + \mu\,  \mathcal{J}_2 (\mathbf{X}) 
%	& = \|\mathbf{Y} - \mathbf{XBS}\|_{F}^{2} + \|\mathbf{Z} - \mathbf{RX}\|_{F}^{2} + \mu\Phi (\bf{X})
\end{aligned}
\end{equation}
where 
\begin{equation}
       \label{eq:J1}
       \mathcal{J}_1 (\mathbf{X}) = \|\mathbf{Y} - \mathbf{XBS}\|_{F}^{2} + \|\mathbf{Z} - \mathbf{RX}\|_{F}^{2} 
\end{equation}
with $\|\cdot\|_{F}$ denoting the matrix Frobenius norm.  $ \mathcal{J}_1 (\mathbf{X}) $ is the data fidelity terms and $ \mathcal{J}_2 (\mathbf{X}) $ is a regularizer that enforces the desired property of the solution while $\mu \ge 0$ is the hyperparameter to balance the data fidelity term $\mathcal{J}_1$ and the regularization term $\mathcal{J}_2$. As shown in~\eqref{eq:object}, prior information on the latent HR HSI is encoded in $ \mathcal{J}_2 (\mathbf{X}) $. Handcrafting a powerful regularizer $ \mathcal{J}_2 (\mathbf{X}) $ is non-trivial. Recently, benefitting from the variable splitting techniques,  plug-and-play methods have been proposed to solve various hyperspectral image inverse problems~\cite{sreehari2016plug, teodoro2018convergent,wang2020learning}. On the contrary, in this work, we construct a regularizer leveraging the network output $\widetilde{\mathbf{X}}\in\mathbb{R}^{B\times N}$ to enforce the solution of~\eqref{eq:J1} to be close to the learned deep priors.  More specifically, let $ \mathcal{J}_2 (\mathbf{X}) $ be the squared Euclidean distance between ${\mathbf{X}}$ and $\widetilde{\mathbf{X}}$:
\begin{equation}
        \label{eq:J2}
        \mathcal{J}_2 (\mathbf{X}) =   \|\mathbf{X} - \widetilde{\mathbf{X}}\|_{F}^{2}
\end{equation}
Then~\eqref{eq:object} is given by
\begin{equation}\label{eq:object2}
\arg\mathop{\min}_{{\bf X}}\, \|\mathbf{Y} -\mathbf{XBS}\|_{F}^{2} + \|\mathbf{Z} - \mathbf{RX}\|_{F}^{2} + \mu\|\mathbf{X} - \widetilde{\mathbf{X}}\|_{F}^{2}
\end{equation}
Compared to other possible formulations, \eqref{eq:object2} allows a straight-forward solver due to the differentiability of the Frobenius norm. Using a powerful priors structure for $\widetilde{\mathbf{X}}$, appropriately integrating the priors into the degradation model as well as balancing contribution of the priors are three key points to obtain a good estimation of the latent HR HSI. 

\section{Proposed Method}
\label{section:proposed_method}
In this section, we present the proposed deep priors-based HSI super-resolution approach in detail. 
Firstly, we introduce the architecture design of  the two-stream fusion network. Then, we elaborate on an efficient approach to the bi-objective optimization problem in~\eqref{eq:object2} based on a Sylvester analytical solver. Finally, we balance the contribution of the learned deep priors by seeking the optimal value of hyperparameter $\mu^*$ under the  {MDC}. The overall scheme of our method is illustrated in Fig.~\ref{fig:scheme}.

\subsection{Deep Priors Learning and Network Design}

Instead of using handcrafted regularizers, we propose to learn the priors of the latent HR HSI from hyperspectral datasets by designing a deep CNN with the aforementioned two-stream architecture, i.e., TSFN. 
%Note that the deep priors learning process is performed in the 3D image domain.

In contrast to the single-stream architecture used in~\cite{dian2018deep, xie2019hyperspectral}, the two-stream deep neural networks  extract and combine features of bimodal data at the feature level, which has been proved to be a powerful tool in applications including videos action recognition~\cite{simonyan2014two, feichtenhofer2016convolutional}, hyperspectral classification~\cite{xu2017multisource} and {hyperspectral superresolution~\cite{li2019joint, deng2020deep,shi2018deep}.} Specifically, we employ residual blocks~\cite{he2016deep} and a skip-connection~\cite{he2016identity} in our network. These techniques have been shown useful in boosting the performance in SISR approaches~\cite{kim2016deeply, ledig2017photo}.

\begin{figure*}[htp]
	\centering
	\includegraphics[scale=1]{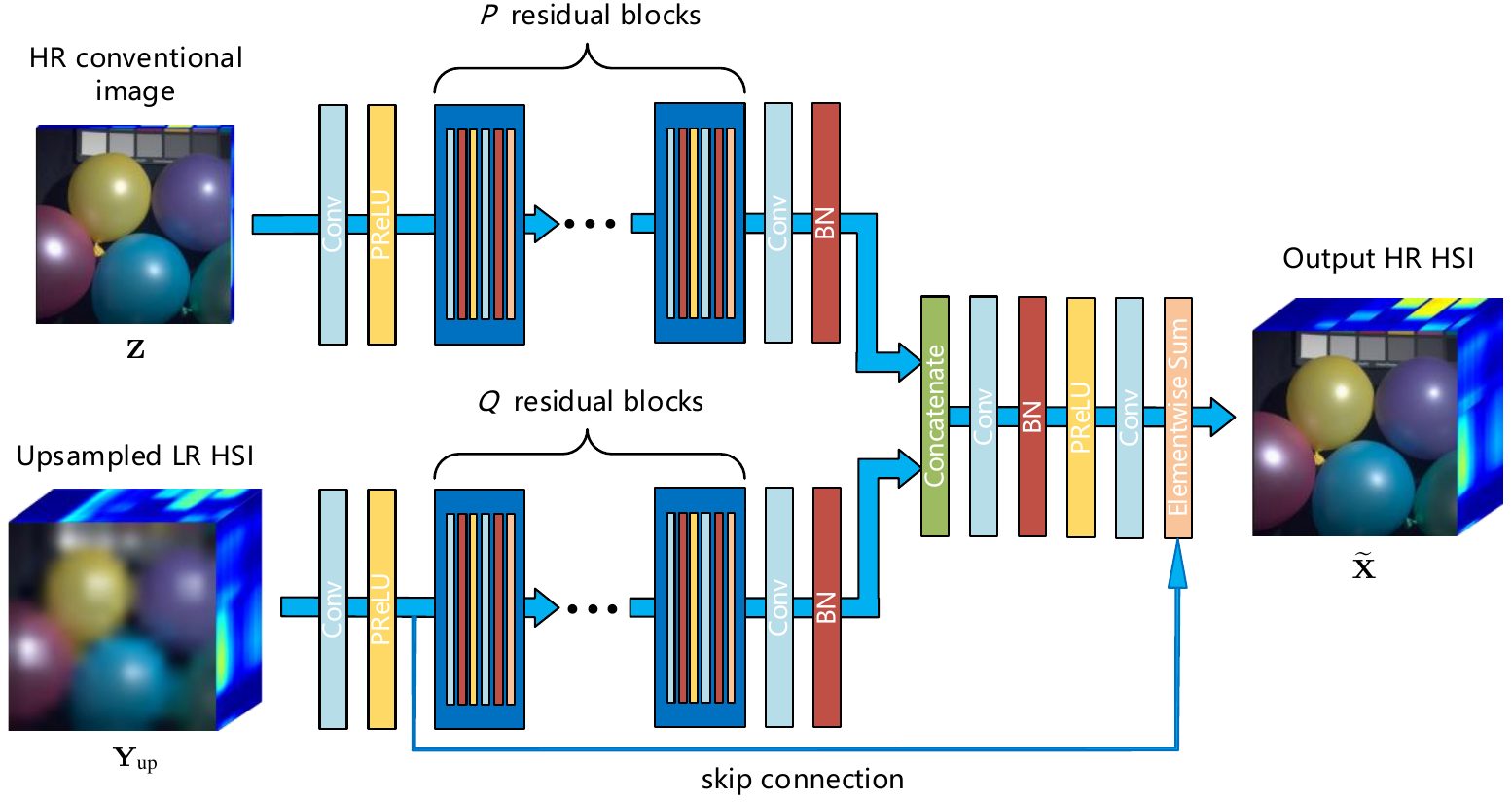}
	\vspace{0mm}
	\caption{Two-stream architecture for hyperspectral image super-resolution.}
	\label{fig:net}
	\vspace{0mm}
\end{figure*}

\begin{figure}[tp]
	\centering
	\includegraphics[scale=0.8]{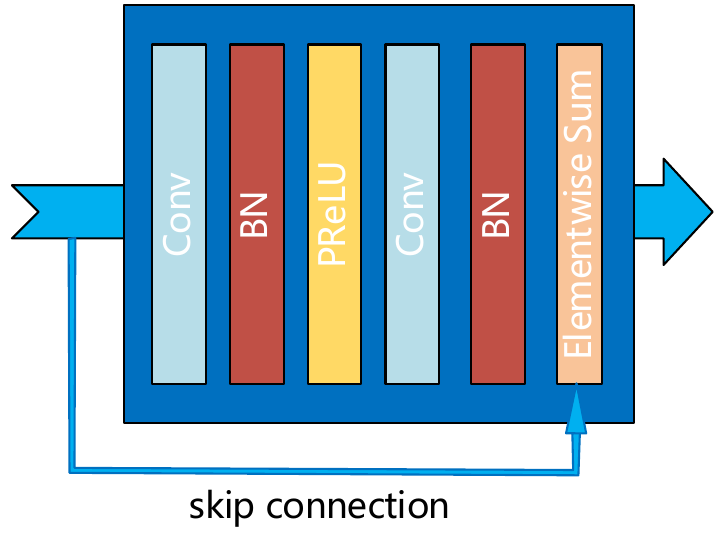}
	\vspace{0mm}
	\caption{Detail components of the residual block.}
	\label{fig:resblock}
	\vspace{0mm}
\end{figure}

As illustrated in Fig.~\ref{fig:net}, the proposed TSFN architecture takes an up-sampled image $\mathbf{Y}_\text{up}\in\mathbb{R}^{B\times N}$ (produced from an LR HSI $\mathbf{Y}$ via the bicubic interpolation) and an HR conventional image $\mathbf{Z}$ as two inputs. 
%In order to achieve a robust HR HSI estimation $\widetilde{\mathbf{X}}$, the spatial context in $\mathbf{Z}$  and spectral attribute in $\mathbf{Y}$ are simultaneously learned.
The feature maps of $\mathbf{Z}$ and $\mathbf{Y}$ extracted by two corresponding streams, containing $P$ and $Q$ residual blocks respectively, are concatenated. Then, the concatenated features are fed into the next convolutional layer.
The core of our network is the residual block, containing two convolutional (Conv) layers followed by batch-normalization (BN) layers~\cite{ioffe2015batch} and ParametricReLU (PReLU)~\cite{he2015delving} as the activation function (see Fig.~\ref{fig:resblock}). The batch-normalization layer is used to speed up the training process as well as boost the image restoration performance. The skip connection passes input feature maps to the output of the residual block via element-wise sum.
A skip connection operator is also employed to add the shallowest feature maps extracted from $\mathbf{Y}_\text{up}$ to the deepest feature maps of the network to reconstruct the final $\widetilde{\mathbf{X}}$. The last convolutional layer contains $B$ filters and the others are composed of $64$ filters, and the kernel size of each filter is $3\times 3$ with a stride of 1. The matrix $\ell_1$-norm based loss function is used:
\begin{equation}\label{eq:loss}
\ell(\Theta) = \sum_{m=1}^{M}  \|\mathcal{F}(\mathbf{Y}_{\text{up},m}, \mathbf{Z}_m;\Theta)-\mathbf{X}_m\|_{1}
\end{equation}
where $\{(\mathbf{Y}_{\text{up},m}, \mathbf{Z}_m;\mathbf{X}_m)\}_{m=1}^{M}$ represent $M$ training image (patch) pairs used to train the network function $\mathcal{F}$ parameterized by~$\Theta$.  {It has been demonstrated that $\ell_{1}$-norm is more appropriate than $\ell_{2}$-norm in low-level image processing tasks as it experimentally provides better performance~\cite{zhao2016loss}. In Subsection~\ref{section:ablation_study}, we conduct related ablation experiments on the loss function in Eq.~\eqref{eq:loss} and give both quantitative and qualitative comparisons to further demonstrate this point. We also show additional ablation studies on the effect of identical residual block numbers $P, Q$ and of the skip connection operator to validate the design choice in our network architecture.}

It is worth noting that the proposed TSFN can be applied to effectively learn priors of the latent HR HSI in degraded scenarios with various scaling factor settings (e.g., 8, 16 and 32). {Our network architecture bears some resemblance to the one in~\cite{shi2018deep}. The main difference is that~\cite{shi2018deep} used the attention residual block while our model extracts features through the residual block without the attention mechanism. Further, we avoid adopting the deconvolution layer and train our model with a mixture of image pairs with different scaling factors. In this way, our trained model can flexibly handle different scenarios.}
\subsection{Integrating Deep Priors into the Degradation Model}
The fusion result inferred from the proposed TSFN, referred to $\widetilde{\mathbf{X}}$, is used in the regularizer $ \mathcal{J}_2$. More specifically, $\widetilde{\mathbf{X}}$, a representation of the prior spatial-spectral information of the latent ${\mathbf{X}}$, is used to regularize the final estimation $\widehat{\mathbf{X}}$, as illustrated in the bi-objective optimization problem~\eqref{eq:object2}. To solve~\eqref{eq:object2}, we force the derivative of the loss function w.r.t. $\mathbf{X}$ to be zero. Thus, the optimum of~\eqref{eq:object2} is defined by the solution of following Sylvester equation:
\begin{equation}\label{eq:Sylvester}
\mathbf{C_1}\widehat{\mathbf{X}} + \widehat{\mathbf{X}}\mathbf{C_2} = \mathbf{C_3}
\end{equation}
where 
\begin{equation}
\begin{split}
	\mathbf{C_1} &= \mathbf{R}^{T}\mathbf{R}+\mu \mathbf{I}_{B}\\
	\mathbf{C_2} &= (\mathbf{BS})(\mathbf{BS})^{T}\\
	\mathbf{C_3} &= \mathbf{R}^{T}\mathbf{Z} + \mathbf{Y}(\mathbf{BS})^{T} + \mu\widetilde{\mathbf{X}}
\end{split}
\end{equation}
where $\mathbf{I}_{B}$ represents the identity matrix of size $B\times B$. According to the well-known conclusion in~\cite{bartels1972solution}, the Sylvester equation in~\eqref{eq:Sylvester} has a unique solution if and only if an arbitrary sum of the eigenvalues of $\mathbf{C_1}$ and $\mathbf{C_2}$ is not equal to zero. Note that $\mathbf{R}^{T}\mathbf{R}$ and $\mathbf{I}_{B}$ are both positive-definite matrices, and thus, $\mathbf{C_1}$ is positive-definite. Further considering that $\mathbf{C_2}$ is positive semi-definite, an arbitrary sum of the eigenvalues of $\mathbf{C_1}$ and $\mathbf{C_2}$ is greater than zero, which ensures the uniqueness of the solution in~\eqref{eq:Sylvester}. 
\renewcommand{\algorithmicrequire}{ \textbf{Input:}} %Use Input in the format of Algorithm
\renewcommand{\algorithmicensure}{ \textbf{Output:}} %UseOutput in the format of Algorithm
\begin{algorithm}[!t]
	\caption{Solution by solving the Sylvester Equation w.r.t. $\widetilde{\mathbf{X}}$.}
	\label{alg_1}
	\begin{algorithmic}
		\REQUIRE $\mathbf{Y}$, $\mathbf{Z}$, $\mathbf{B}$, $\mathbf{S}$, $\mathbf{R}$,  $\widetilde{\mathbf{X}}$, $\mu$.\\
		\ENSURE $\widehat{\mathbf{X}}$.\\
		\STATE Initialize $\mathbf{C_1} = \mathbf{R}^{T}\mathbf{R}+\mu \mathbf{I}_{B}$,\\
		$\qquad \quad \ \ \, \mathbf{C_2} = (\mathbf{BS})(\mathbf{BS})^{T}$,\\
		$\qquad \quad \ \ \, \mathbf{C_3} = \mathbf{R}^{T}\mathbf{Z} + \mathbf{Y}(\mathbf{BS})^{T} + \mu \widetilde{\mathbf{X}}$.
		\STATE (a) Eigen-decomposition of $\mathbf{B}$:\\
		\centerline{$\mathbf{B} = {\mathbf{FDF}^{H}}$}
		\STATE (b) $\overline{\mathbf{D}} = \mathbf{D}(\mathbf{1}_s\otimes \mathbf{I}_n)$
		\STATE (c) Eigen-decomposition of $\mathbf{C_1}$:\\
		\centerline{$\mathbf{C_1} = {\mathbf{Q\Lambda Q}^{-1}}$}
		\STATE (d) $\overline{\mathbf{C_3}} = \mathbf{Q}^{-1}\mathbf{C_3}\mathbf{F}$
		\STATE (e) Compute auxiliary matrix $\overline{\mathbf{X}}$ band by band
		\FOR{$k=1$ to $B$}
		\STATE $\overline{\mathbf{X}}_{k}=\lambda_{k}^{-1}({\overline{\mathbf{C_3}}})_k - \lambda_{k}^{-1}({\overline{\mathbf{C_3}}})_k\overline{\mathbf{D}}(\lambda_{k}s\mathbf{I}_n + \sum\limits_{t=1}^{s} \mathbf{D}_t^2)\overline{\mathbf{D}}^{H}$
		\ENDFOR
		\STATE (f) $\widehat{\mathbf{X}} = \mathbf{Q}\overline{\mathbf{X}}\mathbf{F}^{H}$
	\end{algorithmic}
\end{algorithm}
The fast algorithm for solving~\eqref{eq:Sylvester} can be achieved by referring to our previous work~\cite{wei2015fast}, with detailed steps summarized in Algorithm~\ref{alg_1}.

\subsection{Adjusting the Contribution from Deep Priors}
{Different  {from} previous deep priors-based approaches~\cite{dian2018deep, xie2019hyperspectral},} achieving a good balance between the data fidelity and regularization terms is also considered in our work. Intuitively, if $\mu$ tends toward infinity, the regularization term $\|\mathbf{X} - \widetilde{\mathbf{X}}\|_{F}^{2}$ will be minimized, i.e., the solution will be $\widetilde{\mathbf{X}}$. On the contrary, the solution tends to minimize the data fidelity terms $\|\mathbf{Y} - \mathbf{XBS}\|_{F}^{2} + \|\mathbf{Z} - \mathbf{RX}\|_{F}^{2}$ when $\mu$ tends to zero. In this subsection, we propose to estimate the regularization parameter with the {MDC} that is initially used for {hyperspectral} image deconvolution~\cite{song2016regularization}. We estimate the response curve of~\eqref{eq:object2} that describes the characteristic of the solution set obtained by varying hyperparameter value $\mu$. Then, MDC is {adopted} to the estimated response curve to seek the optimal value of $\mu$ {in the hyperspectral super-resolution problem by using the golden-section search method.}

Problem~\eqref{eq:object2} can also be considered as a bi-objective optimization in form of
\begin{equation}\label{eq:object3}
\begin{aligned}
\widehat{\mathbf{X}} & = \arg\mathop{\min}_{{\bf X}}\, (\mathcal{J}_1({\mathbf{X}}), \mathcal{J}_2({\mathbf{X}}))
\end{aligned}
\end{equation}
\subsubsection{Pareto Front} The definition of $\textit{Pareto front}$ depends on the notion of $\textit{domination}$ defined in~\cite{deb2001multi}. This notion is important in the bi-objective optimization since it gives a criterion of judging a better solution. Let $\widehat{\mathbf{X}}^{(1)}$ and $\widehat{\mathbf{X}}^{(2)}$ be two different solutions of the bi-objective optimization in~\eqref{eq:object3}. When we say $\widehat{\mathbf{X}}^{(1)}$ dominates $\widehat{\mathbf{X}}^{(2)}$, it means that $\mathcal{J}_{i}(\mathbf{X}^{(1)})$ is not larger than $ \mathcal{J}_{i}(\mathbf{X}^{(2)})$ for all $i\in{\{1,2\}}$ and $\mathcal{J}_{j}(\mathbf{X}^{(1)})$ is smaller than $\mathcal{J}_{j}(\mathbf{X}^{(2)})$ for at least one $j\in{\{1,2\}}$:
\begin{equation}\label{eq:dom}
{\widehat{\mathbf{X}}^{(1)} \preceq \widehat{\mathbf{X}}^{(2)}\ \mathrm{iff} \begin{cases}
\mathcal{J}_{i}(\mathbf{X}^{(1)}) \leq \mathcal{J}_{i}(\mathbf{X}^{(2)}),\quad \forall i\in{\{1,2\}}\\
\exists \ i\in{\{1,2\}} \qquad \mathrm{s.t.}\ \mathcal{J}_{i}(\mathbf{X}^{(1)}) < \mathcal{J}_{i}(\mathbf{X}^{(2)})
\end{cases}}
\end{equation}
Apart form that, the solution $\widehat{\mathbf{X}}^{(1)}$ does not dominate $\widehat{\mathbf{X}}^{(2)}$. The solution is said to be non-dominated or $\textit{Pareto optimal}$ for a bi-objective problem if all other solutions in the 
feasible region have higher values for at least one objective. The set containing all the non-dominated solutions is defined as the $\textit{Pareto front}$. It means that any point in the $\textit{Pareto front}$ cannot be said to dominate any other. The shape of $\textit{Pareto front}$ denotes the set of all the achievable trade-offs between two objectives $\mathcal{J}_{1}(\mathbf{X})$ and $\mathcal{J}_{2}(\mathbf{X})$.
\begin{figure}[tp]
	\centering
	\includegraphics[scale=0.8]{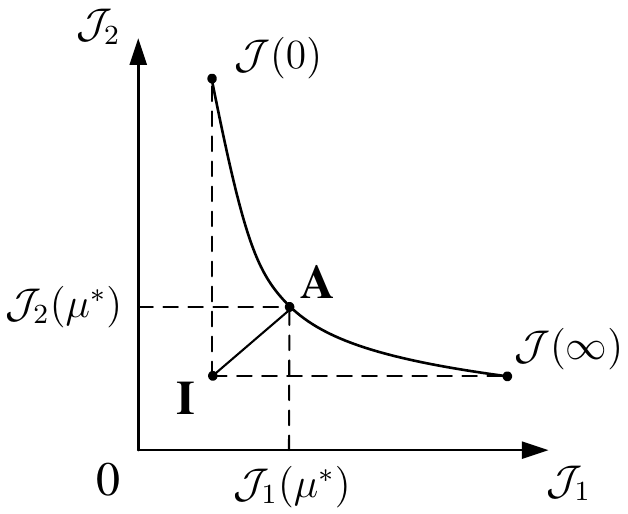}
	\vspace{0mm}
	\caption{ {Representation of the response curve. The optimal point is denoted as $\mathbf{A}\,(\mathcal{J}_1({\mu^*})$, $\mathcal{J}_2({\mu^*}))$.}}
	\label{fig:front}
	\vspace{0mm}
\end{figure}

\begin{figure*}[htp]
	\centering
	\includegraphics[scale=1]{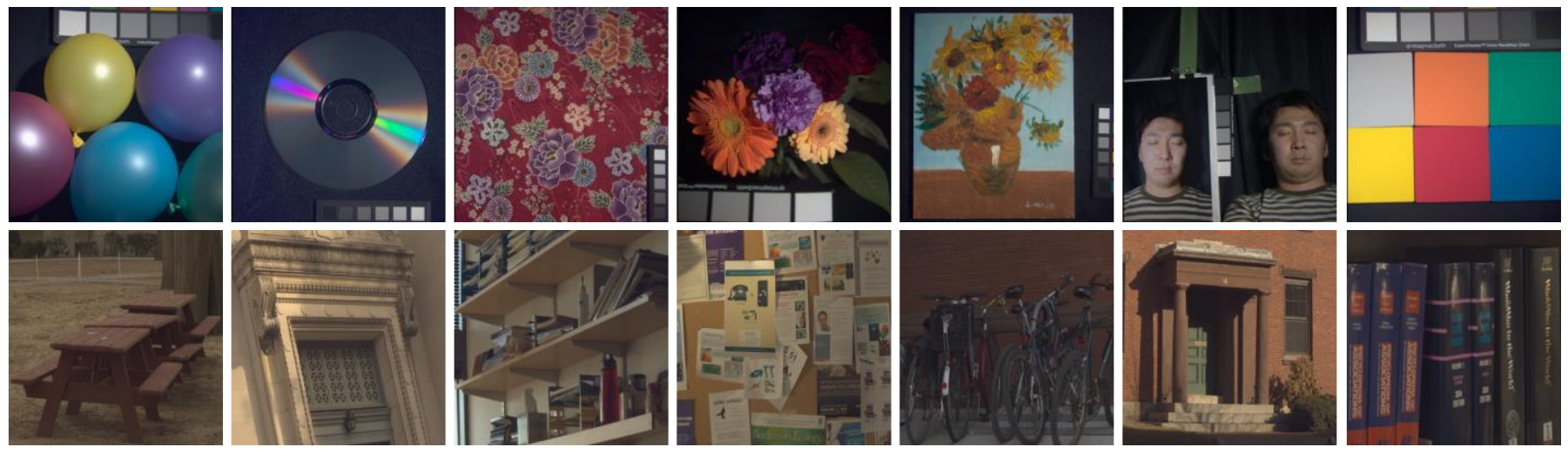}
	\vspace{0mm}
	\caption{Some color images from the CAVE dataset~\cite{yasuma2010generalized} (the first row) and the Harvard dataset~\cite{chakrabarti2011statistics} (the second row).}
	\label{fig:examples}
	\vspace{0mm}
\end{figure*}

In our case, there is a trade-off between $\mathcal{J}_{1}(\mathbf{X})$ and $\mathcal{J}_{2}(\mathbf{X})$ controlled by the regularization parameter $\mu$. By solving~\eqref{eq:object2} via Algorithm \ref{alg_1}, each value of $\mu$ yields a solution:
\begin{equation}\label{eq:xmu}
	\widehat{\mathbf{X}}_{\mu} = \arg\mathop{\min}_{{\bf X}}\, \mathcal{J}(\mathbf{X}) = \arg\mathop{\min}_{{\bf X}}\, \mathcal{J}_1(\mathbf{X}) + \mu \mathcal{J}_2(\mathbf{X})
\end{equation}
and a point $(\mathcal{J}_1(\widehat{\mathbf{X}}_{\mu})$, $\mathcal{J}_2(\widehat{\mathbf{X}}_{\mu}))$ of the response curve. {According to the theorem in~\cite{song2016regularization}, this response curve is convex and exactly coincides with the $\textit{Pareto front}$ because $\mathcal{J}(\mathbf{X})$ is convex. For notation simplicity, we will write the point as $(\mathcal{J}_1({\mu})$, $\mathcal{J}_2({\mu}))$ and $\mathcal{J}(\widehat{\mathbf{X}}_{\mu})$ as $ \mathcal{J}({\mu})$.} 

\renewcommand{\algorithmicrequire}{ \textbf{Input:}} %Use Input in the format of Algorithm
\renewcommand{\algorithmicensure}{ \textbf{Output:}} %UseOutput in the format of Algorithm
\begin{algorithm}[!t]
	\caption{Deep Priors-Based HR Hyperspectral Image Super-Resolution.}
	\label{alg_2}
	\begin{algorithmic}
		\REQUIRE $\mathbf{Y}$, $\mathbf{Z}$, $\mathbf{B}$, $\mathbf{S}$, $\mathbf{R}$, $\Theta$, $\alpha$, $a$, $b$, $\delta$, $\epsilon$.\\
		\ENSURE $\widehat{\mathbf{X}}$.\\
		\STATE Initialize $\mathbf{Y}_\text{up}$ from $\mathbf{Y}$ with the bicubic interpolation.\\
		
		\STATE (a) Obtain representation of learned deep-priors via TSFN:\\
		\centerline{$\widetilde{\mathbf{X}} = \mathcal{F}(\mathbf{Y}_{\text{up}}, \mathbf{Z};\Theta)$}
		\STATE (b) Search the optimal hyperparameter:
		\REPEAT
		\STATE $\mu_{1} = a + \delta(b - a)$;
		\STATE $\mu_{2} = b - \delta(b - a)$;
		\STATE Compute $\widehat{\mathbf{X}}_{\mu_{1}}$ and $\widehat{\mathbf{X}}_{\mu_{2}}$ via~\eqref{eq:object2} using Algorithm \ref{alg_1};
		\STATE Obtain $\mathcal{D}(\mu_{1})$ and $\mathcal{D}(\mu_{2})$ via~\eqref{eq:distance};
		\IF {$\mathcal{D}(\mu_1) < \mathcal{D}(\mu_2)$}
		\STATE $b = \mu_2$
		\ELSE 
		\STATE $a = \mu_1$
		\ENDIF
		\UNTIL {$b - a < \epsilon$}
		\STATE (c) $\mu^{*} = \alpha (a+b) / 2$;
		\STATE (d) Output $\widehat{\mathbf{X}} = \widehat{\mathbf{X}}_{\mu^{*}}$ via~\eqref{eq:object2} using Algorithm \ref{alg_1};
	\end{algorithmic}
\end{algorithm}

\subsubsection{{Minimum Distance Criterion}}
{The minimum distance criterion~\cite{song2016regularization} allows us to search the optimal point $\mathbf{A}\,(\mathcal{J}_1({\mu^*})$, $\mathcal{J}_2({\mu^*}))$ on the response curve with $\mu^{*}$ being the optimal parameter value. As represented in Fig.~\ref{fig:front}, this optimal point  {$\mathbf{A}$} is at minimum distance to the $\textit{ideal point}\ \mathbf{I}$, which is defined as $(I_1, I_2)$ and corresponds to the point whose coordinates are minimum in the two objectives:}
\begin{equation}\label{eq:ideal}
\begin{split}
&I_1 = \mathcal{J}(0) = \arg\mathop{\min}_{{\mathbf{X}}}\, \mathcal{J}_{1}(\mathbf{X})\\
&I_2 = \mathcal{J}(\infty) = \arg\mathop{\min}_{\mathbf{X}}\, \mathcal{J}_{2}(\mathbf{X})
\end{split}
\end{equation}
{Unlike the HSI deconvolution problem in~\cite{song2016regularization}, the value of $\mathcal{J}_1$ is typically much smaller than $\mathcal{J}_2$ in the HSI super-resolution problem since $\mathcal{J}_1$ involves less pixels ($B \times n \times n$ and $b \times N \times N$) compared to $\mathcal{J}_2$ with $B \times N \times N$ pixels. Considering the large scale difference between $\mathcal{J}_1$ and $\mathcal{J}_2$, we propose to use a scaled distance as}
\begin{equation}\label{eq:distance}
	\mathcal{D}(\mu) = (\mathcal{J}_1({\mu}) - {I}_1)^2 + \alpha (\mathcal{J}_2({\mu}) - {I}_2)^2
\end{equation}
{where $\alpha = (b / B) ^ 2 + (1 / s^2)^2$ is the scaling factor. Then the optimal hyperparameter is given by}
\begin{equation}\label{eq:mdc}
\mu^{*} = \alpha \arg\mathop{\min}_{{\mu}}\,\mathcal{D}(\mu).
\end{equation}
The varying range of $\mu$ is $(0, \infty)$. In practice, the value of $\mu$ cannot be set to $\infty$ but can be fixed to a large value {$b$}. Meanwhile, the lower bound of $\mu$ is set to a small value {$a$} rather than zero to avoid leading to highly ill-posed problems. 

\subsubsection{{Optimal Point Search}} {The MDC is proved unimodal and always admits a unique minimum since the response curve is convex~\cite{song2016regularization}. It is possible to design a fast approach aiming at finding the optimal point $\mathbf{A}$ on the response curve. We propose to use the golden-section search method that is able to efficiently find the optimum point for the unimodal function~\cite{kim1997iterated}. This method operates in the interval $[a, b]$ and generates two intermediate points :}
\begin{equation}\label{eq:golden}
\begin{split}
&\mu_1 = a + \delta(b - a)\\
&\mu_2 = b - \delta(b - a)\\
\end{split}
\end{equation}
{where  $\delta =0.618$ is the golden ratio.
The evaluated values of $\mathcal{D}(\mu_1)$ and $\mathcal{D}(\mu_2)$ are then compared and, if $\mathcal{D}(\mu_1) < \mathcal{D}(\mu_2)$, then $\mu_2$ replaces $b$ (else, $\mu_1$ replaces $a$). This procedure is repeated in the new smaller interval $[a, b]$ until $b - a < \epsilon$ where $\epsilon > 0$ is an allowable final length of uncertainty. Finally, the estimated optimal hyperparmeter is given by}
\begin{equation}\label{eq:est_opt}
\mu^{*} = \alpha (a+b) / 2
\end{equation}

The proposed overall HSI super-resolution framework is summarized in Algorithm \ref{alg_2}.

\section{Experiments}
In this section, several experiments are conducted to illustrate the effectiveness of our framework. The results provided by the proposed method are compared with those of state-of-the-art HSI super-resolution methods from both quantitative and qualitative perspectives. 

\begin{figure*}[htp]
	\centering
	\includegraphics[scale=1]{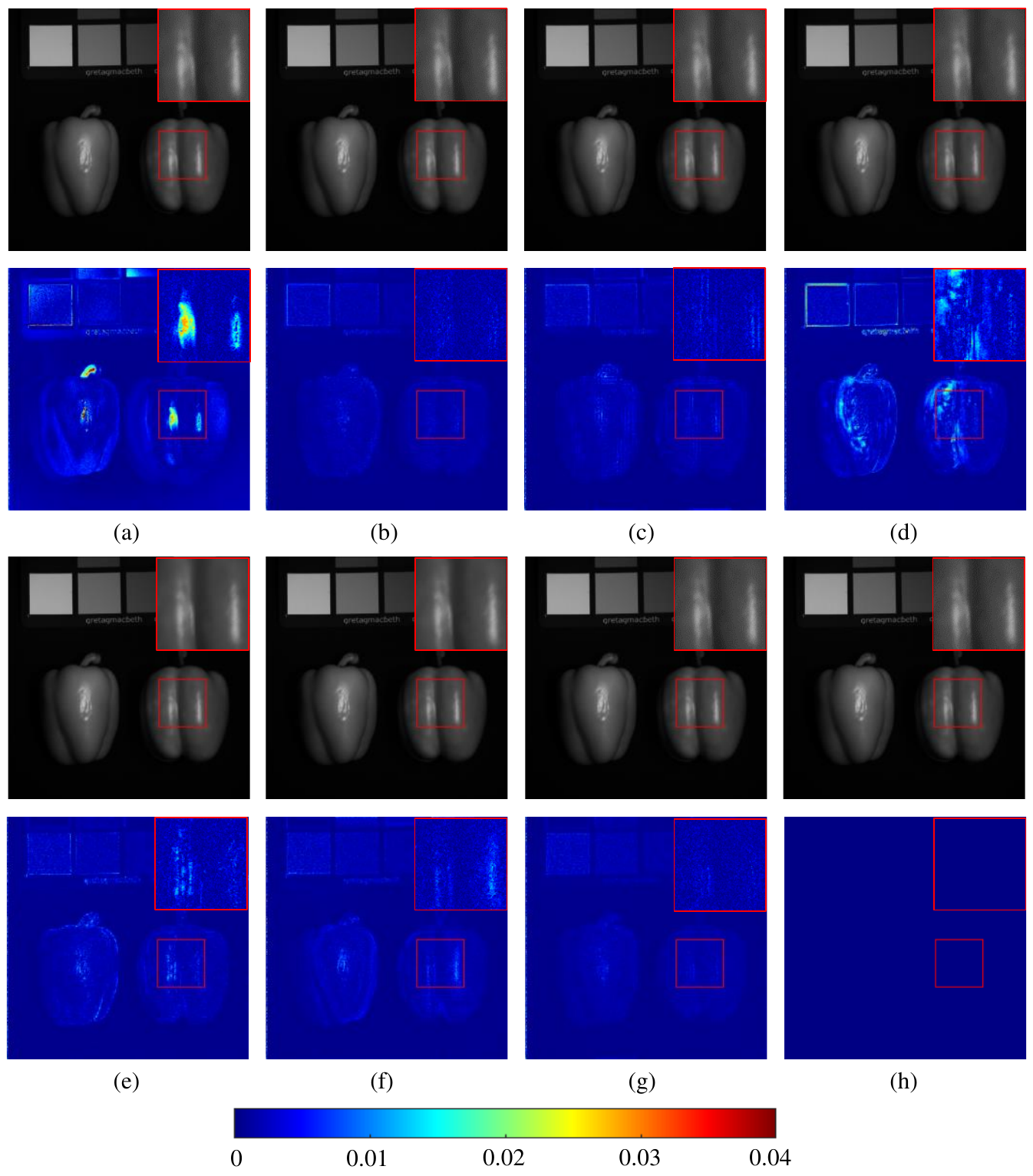}
	\caption{ {Reconstructed images and corresponding error maps of the image \emph{Real and fake peppers} from the CAVE dataset {($s=8$)} at 550 nm band. Methods (RMSE/PSNR): (a) CSU (0.91/ 48.93). (b) NSSR (0.54/53.41). (c) DHSIS (0.58/52.91). (d) CMS (0.86/49.42). (e) LTTR (0.61/52.49). (f) UAL (0.58/52.87). (g) Proposed ($\bf 0.51/54.01$). (h) Ground Truth.}}
	\label{fig:cave}
\end{figure*}
\begin{table*}[]
	\centering
	\caption{ {Average RMSE, PSNR, ERGAS, SAM of different methods with three scaling factors ($s = 8, 16, 32$) on the CAVE dataset.}}
	 {
		\begin{tabular}{c|c|c|c|c|c|c|c|c}
			\hline \hline
			& Metric                                                                           & CSU                       & NSSR                      & DHSIS                     & CMS      & LTTR     & UAL            & Proposed                  \\ \hline \hline
			\multirow{4}{*}{$s=8$}  & 
			\multirow{4}{*}{\begin{tabular}[c]{@{}c@{}}RMSE\\ PSNR\\ ERGAS\\ SAM\end{tabular}}& \multirow{4}{*}{\begin{tabular}[c]{@{}c@{}}2.57\\41.76\\1.196\\6.27\end{tabular}} & 
			\multirow{4}{*}{\begin{tabular}[c]{@{}c@{}}1.47\\46.66\\0.665\\3.72\end{tabular}} & 
			\multirow{4}{*}{\begin{tabular}[c]{@{}c@{}}1.36\\47.01\\0.621\\3.74\end{tabular}} & 
			\multirow{4}{*}{\begin{tabular}[c]{@{}c@{}}1.65\\45.29\\0.734\\3.89\end{tabular}} & 
			\multirow{4}{*}{\begin{tabular}[c]{@{}c@{}}1.49\\46.55\\0.672\\3.84\end{tabular}} & 
			\multirow{4}{*}{\begin{tabular}[c]{@{}c@{}}1.42\\46.47\\0.640\\3.60\end{tabular}} & 
			\multirow{4}{*}{\begin{tabular}[c]{@{}c@{}}\bf 1.12 \\ \textbf{48.63}\\ \bf 0.511\\ \bf3.23\end{tabular}} \\
			&                           &                           &                              &                           &                           &                           &                           &                           \\
			&                           &                           &                              &                           &                           &                           &                           &                           \\
			&                           &                           &                              &                           &                           &                           &                           &                           \\\hline
			\multirow{4}{*}{$s=16$} 	&	
			\multirow{4}{*}{\begin{tabular}[c]{@{}c@{}}RMSE\\ PSNR\\ ERGAS\\ SAM\end{tabular}} & \multirow{4}{*}{\begin{tabular}[c]{@{}c@{}}2.82\\41.01\\0.643\\6.47\end{tabular}} & 
			\multirow{4}{*}{\begin{tabular}[c]{@{}c@{}}1.77\\45.31\\0.398\\4.32\end{tabular}} & 
			\multirow{4}{*}{\begin{tabular}[c]{@{}c@{}}1.79\\44.74\\0.391\\4.57\end{tabular}} & 
			\multirow{4}{*}{\begin{tabular}[c]{@{}c@{}}1.99\\43.94\\0.445\\4.37\end{tabular}} & 
			\multirow{4}{*}{\begin{tabular}[c]{@{}c@{}}1.84\\44.87\\0.415\\4.63\end{tabular}} & 
			\multirow{4}{*}{\begin{tabular}[c]{@{}c@{}}1.57\\45.80\\0.345\\3.84\end{tabular}} & 
			\multirow{4}{*}{\begin{tabular}[c]{@{}c@{}}\bf 1.39\\\textbf{47.02}\\\bf 0.310\\\bf 3.76\end{tabular}} \\
			&                           &                           &                              &                           &                           &                           &                           &                           \\
			&                           &                           &                              &                           &                           &                           &                           &                           \\
			&                           &                           &                              &                           &                           &                           &                           &                           \\ \hline
			\multirow{4}{*}{$s=32$} & 
			\multirow{4}{*}{\begin{tabular}[c]{@{}c@{}}RMSE\\ PSNR\\ ERGAS\\ SAM\end{tabular}} & \multirow{4}{*}{\begin{tabular}[c]{@{}c@{}}3.02\\40.44\\0.336\\6.83\end{tabular}} & 
			\multirow{4}{*}{\begin{tabular}[c]{@{}c@{}}2.24\\43.49\\0.244\\5.22\end{tabular}} & 
			\multirow{4}{*}{\begin{tabular}[c]{@{}c@{}}2.45\\42.34\\0.257\\5.87\end{tabular}} & 
			\multirow{4}{*}{\begin{tabular}[c]{@{}c@{}}2.35\\42.66\\0.257\\5.04\end{tabular}} & 
			\multirow{4}{*}{\begin{tabular}[c]{@{}c@{}}2.28\\43.33\\0.248\\5.46\end{tabular}} & 
			\multirow{4}{*}{\begin{tabular}[c]{@{}c@{}}1.85\\44.66\\0.196\\\textbf{4.33}\end{tabular}} & 
			\multirow{4}{*}{\begin{tabular}[c]{@{}c@{}}\textbf{1.80}\\\textbf{45.07}\\\textbf{ 0.190}\\{4.59}\end{tabular}} \\ 
			&                           &                           &                              &                           &                           &                           &                           &                           \\
			&                           &                           &                              &                           &                           &                           &                           &                           \\
			&                           &                           &                              &                           &                           &                           &                           &                           \\ \hline \hline
	\end{tabular}	}
	\label{tab_cave}
\end{table*}

\subsection{Dataset and Experimental Setup}
\label{sec:dataset}
In this study,  {three simulated hyperspectral datasets, i.e., the CAVE dataset~\cite{yasuma2010generalized}, the Harvard dataset~\cite{chakrabarti2011statistics} and the Chikusei dataset~\cite{yokoya2017hyperspectral}, were used to evaluate the performance of our proposed method.} The CAVE dataset consists of 32 indoor HSI recorded under controlled illuminations, each of which is of size $512\times 512$ in spatial domain and contains 31 spectral bands ranging from 400 nm to 700 nm at a wavelength interval of 10 nm. In the Harvard dataset, there are 50 indoor and outdoor HSI captured under daylight illumination. These images consist of $1392\times 1040$ pixels, with 31 spectral bands of $10$ nm, covering the visible spectrum 420 to 720 nm. The top left $1024\times 1024$ pixels  {were} cropped and extracted  {for} our experiments. Examples of color images from these two datasets are shown in Fig.~\ref{fig:examples}.  {The Chikusei dataset contains an airborne HSI taken by a Visible and Near-Infrared (NIR) imaging sensor over agricultural and urban areas in Chikusei, Ibaraki, Japan. The hyperspectral dataset has 128 bands in the spectral range from 363 nm to 1018 nm and the scene consists of $2517 \times 2335$ pixels. After removing the black boundaries in the spatial domain, centering $2048 \times 2048$ pixels were cropped and extracted for our experiments.}

In our experiments, we  {followed} the standard HSI super-resolution setups as in~\cite{dong2016hyperspectral} and~\cite{zhang2018exploiting}. More specifically, the HSIs from  {three} datasets  {were} scaled into the range of $[0,1]$, then  {served} as the ground truth of $\mathbf{X}$. The LR HSI $\mathbf{Y}$  {was} generated by down-sampling the ground truth over disjointing $s \times s$ blocks, where scaling factor $s$  {was} set to $8, 16$ and $32$. The HR conventional (RGB) image $\mathbf{Z}$  {was} simulated by down-sampling the ground truth in the spectral domain using the spectral response function $\mathbf{R}$, which  {was} derived from the response of a Nikon D700 camera\footnote{https://www.maxmax.com/faq/camera-tech/spectral-response/nikon-d700-study.}.  {For the Chikusei dataset, considering the diversity of hyperspectral sensors, the spectral response function $\mathbf{R}$ of the IKONOS satellite\footnote{https://www.satimagingcorp.com/satellite-sensors/ikonos/} was employed to generate the HR conventional (RGB-NIR) image $\mathbf{Z}$.} As in~\cite{dian2018deep}, we  {selected} the first 20 images from the CAVE dataset as the training set and the rest 12 images as the test set. In the Harvard dataset, the first 30 images  {were} used for training while the other 20 ones  {were} used for testing.  {In the Chikusei dataset, we  {selected} a $1024 \times 2048$-pixel-size image from the top area of the image for training while the remaining part of the image  {were} cropped into 8 non-overlap $512\times512\times128$ as testing data.}

 {Besides, real sample images \emph{Roman Colosseum} acquired by World View-2 (WV-2) were used in these experiments. This dataset contains an LR HSI and an HR RGB image, which are scaled into the range of [0,1]. The HR RGB image is of size $2632\times 1676$ in spatial domain and the LR HSI is of size $658\times 419$ with 8 bands in spectral range from 400 to 1040 nm. Following~\cite{bu2020hyperspectral}, the blurring matrix and the spectral response function were estimated using the method in~\cite{simoes2014convex}. The ground-truth HR HSI in real data is not available so that training image (patch) pairs $\{(\mathbf{Y}_{\text{up},m}, \mathbf{Z}_m;\mathbf{X}_m)\}_{m=1}^{M}$ were used to be generated for training the network function. To tackle this issue, we considered the LR HSI as the ground truth $\mathbf{X}$ and down-sampled it to generate network inputs $\mathbf{Y_{\text{up}}}$ and $\mathbf{Z}$ using the estimated blurring matrix and the spectral response function.}

%\begin{figure*}[htp]
%	\centering
%	\includegraphics[scale=0.90]{Figures/spec_diff_2.pdf}
%	\vspace{0mm}
%	\caption{Spectral differences with respect to the ground truths at three randomly selected locations. (a)-(c): image \emph{Real and fake peppers} from the CAVE dataset. (d)-(f): image \emph{imgf1} from the Harvard dataset {when the scaling factor $s = 16$}.}
%	\label{fig:spec_diff}
%	\vspace{0mm}
%\end{figure*}

\subsection{ {Implementation Details}}
We  {implemented} the proposed two-stream fusion network using PyTorch framework and  {initialized} the model using the method $\textit{HeUniform}$~\cite{he2015delving}. The Adam optimizer~\cite{kingma2014adam}  {was} utilized to minimize the loss function in~\eqref{eq:loss} with an initial learning rate 0.0002 and a mini-batch of 16 in 500 epochs.  {In the training phase, each original HR HSI was cropped into 100, 400, 800 patches of size 128$\times$128 for the CAVE, Harvard, Chikusei dataset, respectively.} Each patch  {was} randomly flipped and rotated for data augmentation before {randomly} generating an LR HSI and an HR conventional image {in different degraded scenarios.} When testing, we  {removed} batch-normalization~\cite{ioffe2015batch}.  {Each stream contains 6 identical residual blocks ($P$,$\,Q\,$=$\,$6) in our model.} {Note that the proposed two-stream fusion network  {was} trained separately on  {four} datasets.}
 {For the golden-section search method, we set $a = 10^{-8}$, $b = 1$ and $\epsilon = 0.01$.}
\begin{figure}[tp]
	\centering
	\includegraphics[scale=0.79]{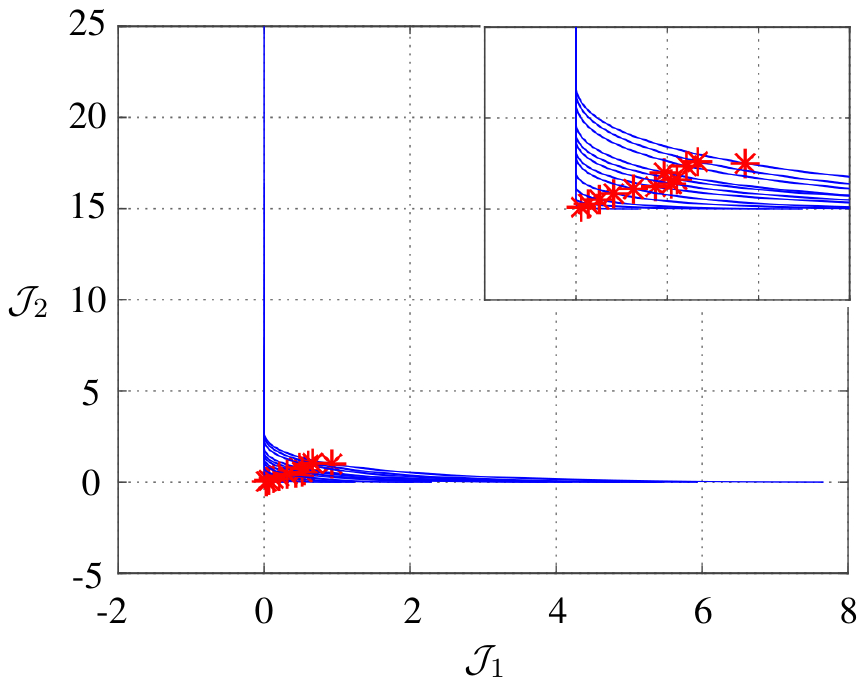}
	\vspace{0mm}
	\caption{ {Response curves (blue) and optimal points (red) with the scaling factor $s = 8$ on the CAVE dataset. Each curve corresponds to a test image.}}
	\label{fig:curve}
	\vspace{0mm}
\end{figure} 
\subsection{Compared Methods and Quantitative Metrics}
We compared the results of our method with  {six} state-of-the-art HSI super-resolution methods: Coupled Spectral Unmixing (CSU) method~\cite{lanaras2015hyperspectral}, Non-negative Structured Sparse Representation (NSSR) method~\cite{dong2016hyperspectral}, Deep Hyperspectral Image Sharpening (DHSIS) method~\cite{dian2018deep}, Clustering Manifold Structure (CMS) method~\cite{zhang2018exploiting},  {Low Tensor-Train Rank (LTTR) method~\cite{dian2019learning} and Unsupervised Adaption Learning (UAL) method~\cite{zhang2020unsupervised}.}
CSU aims to regularize the HSI super-resolution problem by considering the spectral unmixing constraints. NSSR focuses on the sparse representation of the latent HR HSI as the prior structure for super-resolution. In DHSIS, the priors of latent HR HSI are learned by a deep neural network using massive data pairs {without the use of pre-trained models}. CMS exploits the manifold structure to capture the spatial correlation of the latent HR HSI to constrain the super-resolution scheme.  {LTTR learns correlations among the spatial, spectral and nonlocal modes of the nonlocal similar HR HSI cubes. UAL learns the deep priors and estimate the unknown spatial degradation in a unsupervised manner. Among these compared methods, DHSIS is the closest one to ours. The main difference is that it learned deep priors from a single pre-processed image while we directly learned the prior from two observed images.} All these methods  {were} implemented with their published codes online. {Note that we separately  {trained} the deep neural network for different scaling factors, blurring kernels and datasets in DHSIS.}
To evaluate the quality of reconstructed hyperspectral images, four quantitative metrics including the root-mean-square error (RMSE), the peak-signal-to-noise-ratio (PSNR), the erreur relative globale adimensionnelle de synth\`{e}se (ERGAS)~\cite{wald2000quality} and the spectral angle mapper (SAM)~\cite{yuhas1992discrimination}  {were} used.
%\subsection{Optimal Hyperparameter Value}
\begin{figure}[tp]
	\centering
	\includegraphics[scale=0.8]{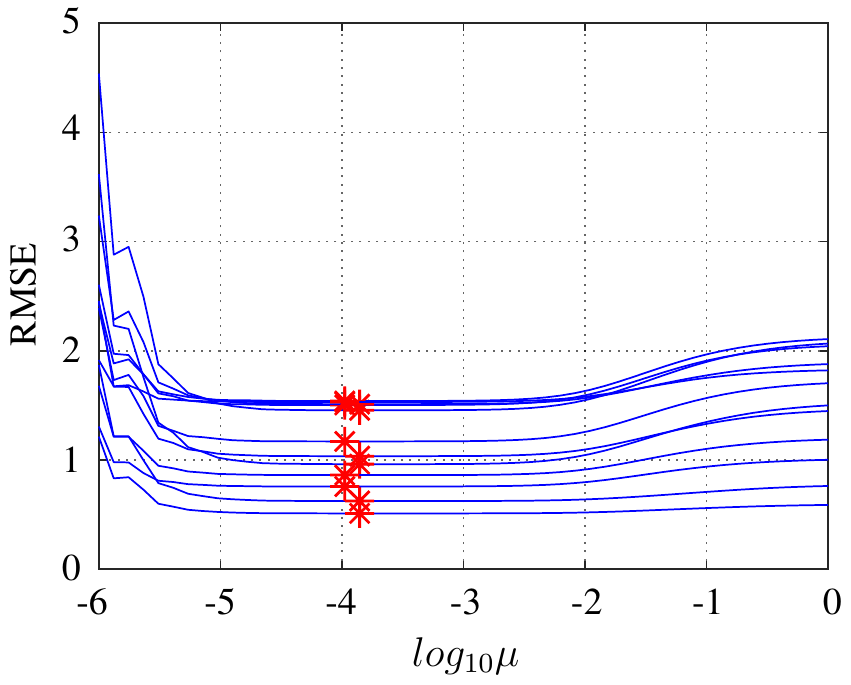}
	\vspace{0mm}
	\caption{ {RMSE curves (blue) and estimated points (red) with the scaling factor $s = 8$ on the CAVE dataset. Each curve corresponds to a test image.}}
	\label{fig:performance}
	\vspace{0mm}
\end{figure} 
\begin{figure*}[htp]
	\centering
	\includegraphics[scale=1]{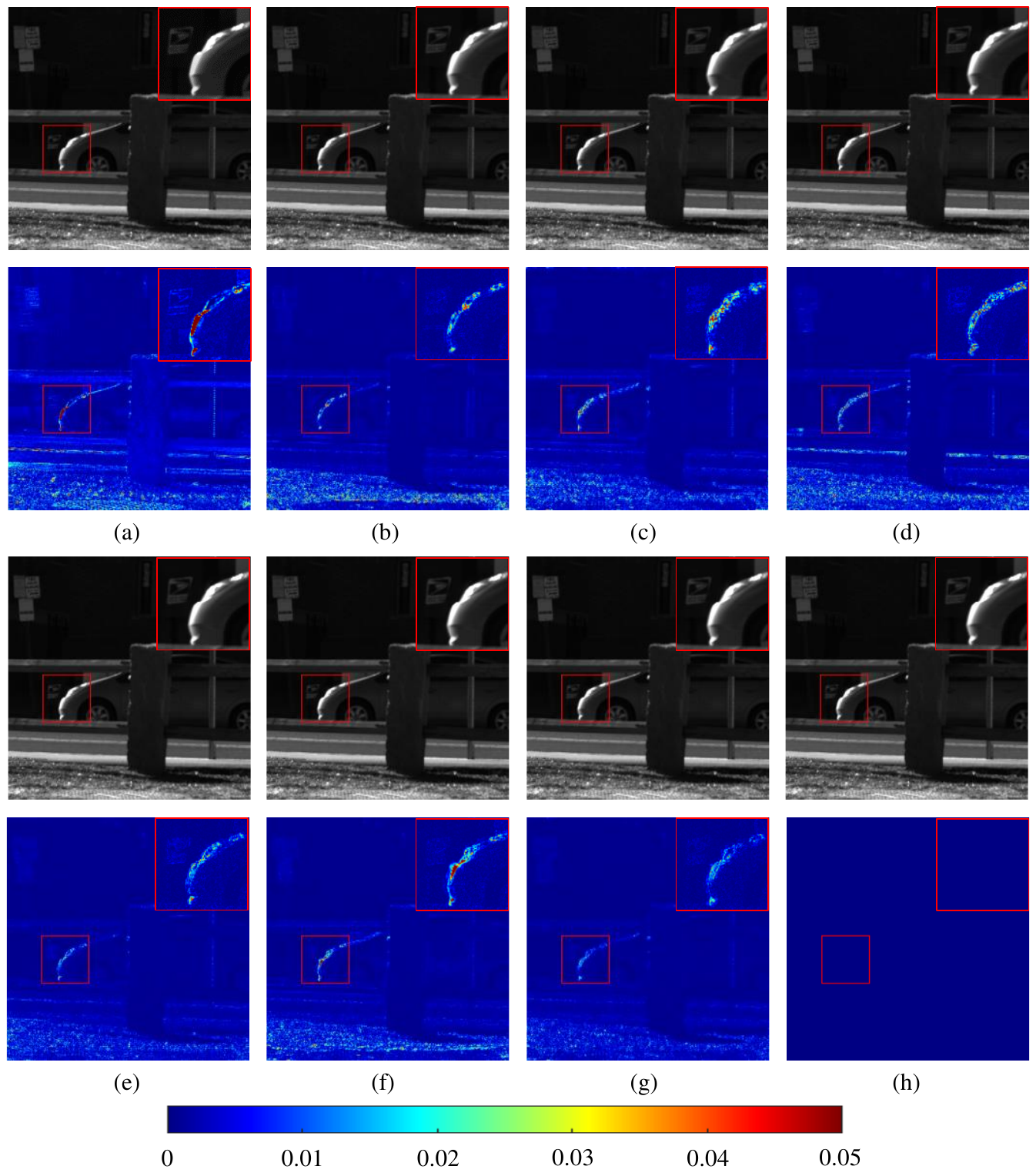}
	\caption{ {Reconstructed images and corresponding error maps of the image \emph{imgf1} from the Harvard dataset {($s=8$)} at 650 nm band. Methods (RMSE/PSNR): (a) CSU (1.49/ 44.69). (b) NSSR (1.12/47.13). (c) DHSIS (0.94/48.70). (d) CMS (1.15/46.94). (e) LTTR (0.82/49.91). (f) UAL (0.91/48.90). (g) Proposed ($\bf 0.73/50.89$). (h) Ground Truth.}}
	\label{fig:harvard}
\end{figure*}
\begin{table*}[]
	\centering
	\caption{ {Average RMSE, PSNR, ERGAS, SAM of different methods with three scaling factors ($s = 8, 16$ and $32$) on the Harvard dataset.}}
	 {
	\begin{tabular}{c|c|c|c|c|c|c|c|c}
		\hline \hline
		& Metric                                                                           & CSU                       & NSSR                      & DHSIS                     & CMS          & LTTR     & UAL            & Proposed                  \\ \hline \hline
		\multirow{4}{*}{$s=8$}  & \multirow{4}{*}{\begin{tabular}[c]{@{}c@{}}RMSE\\ PSNR\\ ERGAS\\ SAM\end{tabular}}& 		\multirow{4}{*}{\begin{tabular}[c]{@{}c@{}}1.91\\45.10\\1.423\\3.65\end{tabular}} & 
		\multirow{4}{*}{\begin{tabular}[c]{@{}c@{}}1.67\\46.29\\1.219\\3.46\end{tabular}} & 
		\multirow{4}{*}{\begin{tabular}[c]{@{}c@{}}1.80\\45.63\\1.415\\3.69\end{tabular}} & 
		\multirow{4}{*}{\begin{tabular}[c]{@{}c@{}}1.65\\46.49\\1.355\\3.57\end{tabular}} & 
		\multirow{4}{*}{\begin{tabular}[c]{@{}c@{}}1.69\\46.10\\1.319\\3.52\end{tabular}} & 
		\multirow{4}{*}{\begin{tabular}[c]{@{}c@{}}1.73\\46.16\\1.270\\3.46 \end{tabular}} & 
		\multirow{4}{*}{\begin{tabular}[c]{@{}c@{}}\bf 1.60\\ \bf 46.73\\ \bf 1.127\\ \textbf{3.36}\end{tabular}} \\
		&                           &                           &                              &                           &                           &                           &                           &                           \\
		&                           &                           &                              &                           &                           &                           &                           &                           \\
		&                           &                           &                              &                           &                           &                           &                           &                           \\ \hline
		\multirow{4}{*}{$s=16$} 	&	\multirow{4}{*}{\begin{tabular}[c]{@{}c@{}}RMSE\\ PSNR\\ ERGAS\\ SAM\end{tabular}} & 		 \multirow{4}{*}{\begin{tabular}[c]{@{}c@{}}2.01\\44.59\\0.739\\3.73\end{tabular}} & \multirow{4}{*}{\begin{tabular}[c]{@{}c@{}}1.78\\45.95\\0.653\\3.58\end{tabular}} & \multirow{4}{*}{\begin{tabular}[c]{@{}c@{}}1.80\\45.88\\0.654\\3.61\end{tabular}} & \multirow{4}{*}{\begin{tabular}[c]{@{}c@{}}1.76\\46.06\\0.721\\3.68\end{tabular}} & 
		\multirow{4}{*}{\begin{tabular}[c]{@{}c@{}}1.81\\45.66\\0.709\\3.65\end{tabular}} & 
		\multirow{4}{*}{\begin{tabular}[c]{@{}c@{}}1.77\\46.03\\0.633\\3.50\end{tabular}} & 
		\multirow{4}{*}{\begin{tabular}[c]{@{}c@{}}\textbf{1.72}\\ \textbf{46.32}\\\bf 0.601\\ \textbf{ 3.48}\end{tabular}} \\
		&                           &                           &                              &                           &                           &                           &                           &                           \\
		&                           &                           &                              &                           &                           &                           &                           &                           \\
		&                           &                           &                              &                           &                           &                           &                           &                           \\ \hline
		\multirow{4}{*}{$s=32$} & \multirow{4}{*}{\begin{tabular}[c]{@{}c@{}}RMSE\\ PSNR\\ ERGAS\\ SAM\end{tabular}} & 		 \multirow{4}{*}{\begin{tabular}[c]{@{}c@{}}2.14\\43.91\\0.394\\3.80\end{tabular}} & 
		\multirow{4}{*}{\begin{tabular}[c]{@{}c@{}}1.87\\45.54\\0.363\\3.73\end{tabular}} & 
		\multirow{4}{*}{\begin{tabular}[c]{@{}c@{}}1.92\\45.41\\0.339\\3.81\end{tabular}} & \multirow{4}{*}{\begin{tabular}[c]{@{}c@{}}1.83\\45.79\\0.381\\3.76\end{tabular}} & 
		\multirow{4}{*}{\begin{tabular}[c]{@{}c@{}}1.91\\45.26\\0.375\\3.81\end{tabular}} & 
		\multirow{4}{*}{\begin{tabular}[c]{@{}c@{}}1.83\\45.81\\0.323\\3.58\end{tabular}} & \multirow{4}{*}{\begin{tabular}[c]{@{}c@{}}\textbf{1.80}\\ \textbf{45.99}\\ \bf 0.313\\ \textbf{3.59}\end{tabular}} \\
		&                           &                           &                              &                           &                           &                           &                           &                           \\
		&                           &                           &                              &                           &                           &                           &                           &                           \\
		&                           &                           &                              &                           &                           &                           &                           &                           \\ \hline \hline
	\end{tabular}	}
	\label{tab_harvard}
\end{table*}

\subsection{ {Performance Evaluation on Simulated Data}}
We  {validated} the proposed method to show its effectiveness in terms of super-resolution performance over other compared methods on  {three public} datasets. 
\begin{figure*}[htp]
	\centering
	\includegraphics[scale=1]{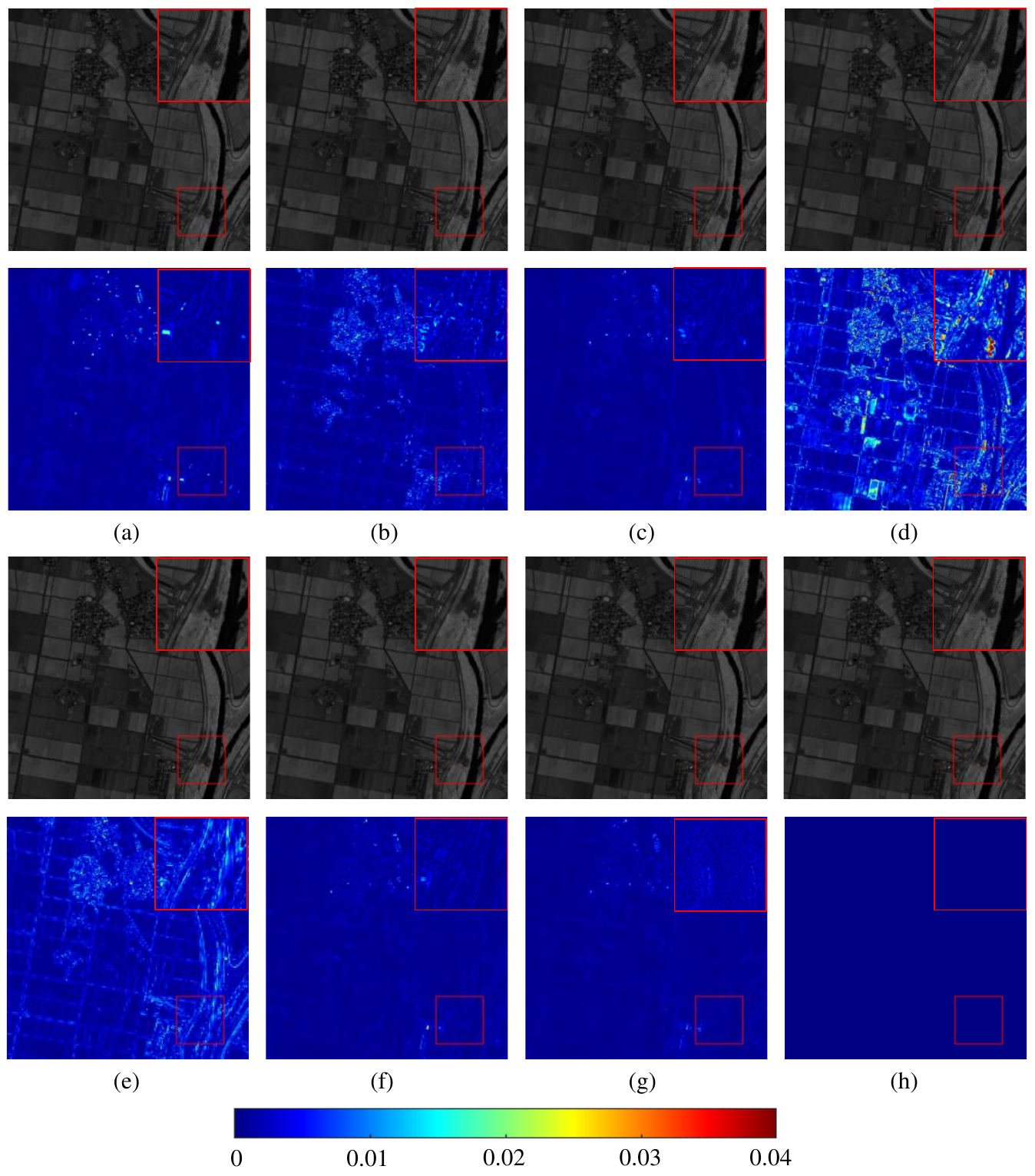}
	\caption{ {Reconstructed images and corresponding error maps of a test image from the Chikusei dataset {($s=8$)} at 800 nm band. Methods (RMSE/PSNR): (a) CSU (0.75/ 50.66). (b) NSSR (0.75/50.59). (c) DHSIS (0.46/54.89). (d) CMS (1.69/43.59). (e) LTTR (0.97/48.38). (f) UAL (0.47/54.73). (g) Proposed ($\bf 0.44/55.35$). (h) Ground Truth.}}
	\label{fig:chikusei}
\end{figure*}
\begin{table*}[htp]
	\centering
	\caption{ {Average RMSE, PSNR, ERGAS, SAM of different methods with three scaling factors ($s = 8, 16$ and $32$) on the Chikusei dataset.}}
	 {
		\begin{tabular}{c|c|c|c|c|c|c|c|c}
			\hline \hline
			& Metric                                                                           & CSU                       & NSSR                      & DHSIS                     & CMS          & LTTR     & UAL            & Proposed                  \\ \hline \hline
			\multirow{4}{*}{$s=8$}  & \multirow{4}{*}{\begin{tabular}[c]{@{}c@{}}RMSE\\ PSNR\\ ERGAS\\ SAM\end{tabular}}& 		\multirow{4}{*}{\begin{tabular}[c]{@{}c@{}}0.89\\51.16\\1.409\\1.79\end{tabular}} & 
			\multirow{4}{*}{\begin{tabular}[c]{@{}c@{}}1.36\\49.90\\1.457\\2.36\end{tabular}} & 
			\multirow{4}{*}{\begin{tabular}[c]{@{}c@{}}0.72\\53.16\\1.181\\1.58\end{tabular}} & 
			\multirow{4}{*}{\begin{tabular}[c]{@{}c@{}}2.42\\44.64\\1.904\\2.97\end{tabular}} & 
			\multirow{4}{*}{\begin{tabular}[c]{@{}c@{}}2.09\\45.67\\2.072\\3.33\end{tabular}} & 
			\multirow{4}{*}{\begin{tabular}[c]{@{}c@{}}0.85\\52.47\\1.474\\1.85\end{tabular}} & 
			\multirow{4}{*}{\begin{tabular}[c]{@{}c@{}}\bf 0.66\\\bf 54.02\\\bf 1.128\\\bf 1.43\end{tabular}} \\
			&                           &                           &                              &                           &                           &                           &                           &                           \\
			&                           &                           &                              &                           &                           &                           &                           &                           \\
			&                           &                           &                              &                           &                           &                           &                           &                           \\ \hline
			\multirow{4}{*}{$s=16$} 	&	\multirow{4}{*}{\begin{tabular}[c]{@{}c@{}}RMSE\\ PSNR\\ ERGAS\\ SAM\end{tabular}} & 		 \multirow{4}{*}{\begin{tabular}[c]{@{}c@{}}0.98\\50.32\\0.757\\1.91\end{tabular}} & \multirow{4}{*}{\begin{tabular}[c]{@{}c@{}}1.47\\49.27\\0.793\\2.55\end{tabular}} & \multirow{4}{*}{\begin{tabular}[c]{@{}c@{}}0.78\\52.64\\0.620\\1.71\end{tabular}} & \multirow{4}{*}{\begin{tabular}[c]{@{}c@{}}2.72\\43.80\\1.040\\3.40\end{tabular}} & 
			\multirow{4}{*}{\begin{tabular}[c]{@{}c@{}}2.58\\44.18\\1.195\\4.15\end{tabular}} & 
			\multirow{4}{*}{\begin{tabular}[c]{@{}c@{}}0.73\\53.10\\0.610\\1.58\end{tabular}} & 
			\multirow{4}{*}{\begin{tabular}[c]{@{}c@{}}\bf0.72\\\bf 53.41\\\bf 0.607\\\bf 1.55\end{tabular}} \\
			&                           &                           &                              &                           &                           &                           &                           &                             \\
			&                           &                           &                              &                           &                           &                           &                           &                           \\
			&                           &                           &                              &                           &                           &                           &                           &                           \\ \hline
			\multirow{4}{*}{$s=32$} & \multirow{4}{*}{\begin{tabular}[c]{@{}c@{}}RMSE\\ PSNR\\ ERGAS\\ SAM\end{tabular}} & 		 \multirow{4}{*}{\begin{tabular}[c]{@{}c@{}}1.11\\48.85\\0.411\\2.07\end{tabular}} & 
			\multirow{4}{*}{\begin{tabular}[c]{@{}c@{}}1.64\\48.26\\0.434\\2.89\end{tabular}} & 
			\multirow{4}{*}{\begin{tabular}[c]{@{}c@{}}0.84\\51.96\\0.332\\1.83\end{tabular}} & \multirow{4}{*}{\begin{tabular}[c]{@{}c@{}}3.13\\42.77\\0.583\\4.17\end{tabular}} & 
			\multirow{4}{*}{\begin{tabular}[c]{@{}c@{}}3.41\\42.30\\0.710\\5.40\end{tabular}} & 
			\multirow{4}{*}{\begin{tabular}[c]{@{}c@{}}\bf0.78\\52.74\\\bf0.314\\\bf1.66\end{tabular}} & \multirow{4}{*}{\begin{tabular}[c]{@{}c@{}}\bf0.78\\\bf52.82\\0.322\\1.67\end{tabular}} \\
			&                           &                           &                              &                           &                           &                           &                           &                           \\
			&                           &                           &                              &                           &                           &                           &                           &                           \\
			&                           &                           &                              &                           &                           &                           &                           &                           \\ \hline \hline
	\end{tabular}	}
	\label{tab_chikusei}
\end{table*}
\subsubsection{Evaluation on CAVE Dataset}
Firstly, we  {conducted and compared} all methods to restore each latent HR HSI $\mathbf{X}$ from the corresponding LR HSI $\mathbf{Y}$ and HR conventional image $\mathbf{Z}$ with scale factor $s$ in the CAVE dataset. The average values of RMSE, PSNR, ERGAS and SAM of the compared methods with different scaling factors on the CAVE dataset are reported in Table \ref{tab_cave}. It is  clear that the proposed method outperforms all competing methods for all different settings. Especially when the scaling factor is small (e.g., $s = $ 8), the superiority and robustness of our method are more significant. The improvement mainly stems from the effectiveness of the learned priors considering both spatial context and spectral attributes of the latent HR HSIs. Furthermore, the regularization parameter estimation also contributes to the satisfactory performance by giving a good trade-off between the data fidelity and regularization terms. Note that the regularization parameter of all competing methods have been fixed for all tested images.
 
 \begin{table*}[htp]
 	\centering
 	\caption{ {Average RMSE, PSNR, ERGAS, SAM of different methods with the scaling factor $s = 8$ (Gaussian blur kernel) on the CAVE, Harvard and Chikusei datasets.}}
 	 {
 		\begin{tabular}{c|c|c|c|c|c|c|c|c}
 			\hline \hline
 			& Metric                                                                           & CSU                       & NSSR                      & DHSIS                     & CMS        & LTTR     & UAL                   & Proposed                  \\ \hline \hline
 			\multirow{4}{*}{CAVE}  	&	\multirow{4}{*}{\begin{tabular}[c]{@{}c@{}}RMSE\\ PSNR\\ ERGAS\\ SAM\end{tabular}} & \multirow{4}{*}{\begin{tabular}[c]{@{}c@{}}2.55\\ 41.85\\ 1.138\\ 6.43\end{tabular}} & 
 			\multirow{4}{*}{\begin{tabular}[c]{@{}c@{}}1.88\\43.82\\0.838\\4.07\end{tabular}} & 
 			\multirow{4}{*}{\begin{tabular}[c]{@{}c@{}}1.36\\46.95\\0.619\\3.75\end{tabular}} & 
 			\multirow{4}{*}{\begin{tabular}[c]{@{}c@{}}1.57\\45.88\\0.686\\3.75\end{tabular}} &
 			\multirow{4}{*}{\begin{tabular}[c]{@{}c@{}}1.51\\46.46\\0.679\\3.85\end{tabular}} &
 			\multirow{4}{*}{\begin{tabular}[c]{@{}c@{}}1.41\\46.53\\0.636\\3.58\end{tabular}} & 
 			\multirow{4}{*}{\begin{tabular}[c]{@{}c@{}}\bf 1.13\\\textbf{48.56}\\\textbf{0.512}\\\bf3.22\end{tabular}} \\
 			&                           &                           &                            &                           &                           &                           &                           &                           \\
 			&                           &                           &                            &                           &                           &                           &                           &                           \\
 			&                           &                           &                            &                           &                           &                           &                           &                           \\ \hline
 			\multirow{4}{*}{Harvard} 	&	\multirow{4}{*}{\begin{tabular}[c]{@{}c@{}}RMSE\\ PSNR\\ ERGAS\\ SAM\end{tabular}} & 		 		
 			\multirow{4}{*}{\begin{tabular}[c]{@{}c@{}}1.91\\45.07\\1.420\\3.72\end{tabular}} & 
 			\multirow{4}{*}{\begin{tabular}[c]{@{}c@{}}1.85\\45.29\\1.275\\3.50\end{tabular}} & 
 			\multirow{4}{*}{\begin{tabular}[c]{@{}c@{}}1.70\\46.29\\1.228\\3.45\end{tabular}} & 
 			\multirow{4}{*}{\begin{tabular}[c]{@{}c@{}}1.63\\46.64\\1.306\\3.54\end{tabular}} & 
 			\multirow{4}{*}{\begin{tabular}[c]{@{}c@{}}1.69\\46.12\\1.306\\3.51\end{tabular}} & 
 			\multirow{4}{*}{\begin{tabular}[c]{@{}c@{}}1.87\\45.52\\1.722\\3.73\end{tabular}} & 
 			\multirow{4}{*}{\begin{tabular}[c]{@{}c@{}}\bf1.58\\\textbf{46.77}\\\bf 1.120\\\textbf{3.34}\end{tabular}} \\
 			&                           &                           &                            &                           &                           &                           &                           &                           \\
 			&                           &                           &                            &                           &                           &                           &                           &                           \\
 			&                           &                           &                            &                           &                           &                           &                           &                           \\ \hline
 			\multirow{4}{*}{Chikusei} 	&	\multirow{4}{*}{\begin{tabular}[c]{@{}c@{}}RMSE\\ PSNR\\ ERGAS\\ SAM\end{tabular}} & 		 		
 			\multirow{4}{*}{\begin{tabular}[c]{@{}c@{}}0.88\\51.04\\1.422\\1.82 \end{tabular}} & 
 			\multirow{4}{*}{\begin{tabular}[c]{@{}c@{}}1.52\\47.86\\1.562\\2.59\end{tabular}} & 
 			\multirow{4}{*}{\begin{tabular}[c]{@{}c@{}}0.73\\52.93\\1.216\\1.58\end{tabular}} & 
 			\multirow{4}{*}{\begin{tabular}[c]{@{}c@{}}2.19\\46.00\\1.839\\3.23\end{tabular}} & 
 			\multirow{4}{*}{\begin{tabular}[c]{@{}c@{}}2.10\\46.65\\2.086\\3.33\end{tabular}} & 
 			\multirow{4}{*}{\begin{tabular}[c]{@{}c@{}}1.26\\50.53\\2.573\\2.73\end{tabular}} & 
 			\multirow{4}{*}{\begin{tabular}[c]{@{}c@{}}\bf 0.66\\\bf 53.97\\\bf 1.124\\\bf 1.43\end{tabular}} \\
 			&                           &                           &                            &                           &                           &                           &                           &                           \\
 			&                           &                           &                            &                           &                           &                           &                           &                           \\
 			&                           &                           &                            &                           &                           &                           &                           &                           \\ \hline \hline
 	\end{tabular}	}
 	\label{tab_gaussian}
 \end{table*}
 
 \begin{table}[tp]
 	\centering
 	\caption{ {Average RMSE, PSNR, ERGAS, SAM of different methods with the scaling factor $s = 8$ (Gaussian blur kernel) on the CAVE dataset under noise conditions.}} 
 	\vspace{2mm}
 	 {	\begin{tabular}{c|cccc}
 			\hline \hline
 			Method       &  RMSE                 &   PSNR         & ERGAS       &      SAM              \\ \hline \hline
 			\multirow{7}{*}{\begin{tabular}[c]{@{}c@{}} CSU \\ NSSR\\ DHSIS \\ CMS \\ LTTR \\ UAL \\ Proposed\end{tabular}} &
 			\multirow{7}{*}{\begin{tabular}[c]{@{}c@{}} 2.68 \\ 2.09 \\ 1.66 \\ 1.78 \\ 1.70 \\ 1.51 \\\bf 1.34 \end{tabular}} & \multirow{7}{*}{\begin{tabular}[c]{@{}c@{}} 41.01 \\42.61 \\ 44.65 \\ 44.30 \\ 44.90 \\ 45.72 \\ \bf 46.42 \end{tabular}} & \multirow{7}{*}{\begin{tabular}[c]{@{}c@{}} 1.231\\0.922\\ 0.741 \\ 0.767 \\ 0.749 \\ 0.673 \\ \bf 0.604 \end{tabular}} & 
 			\multirow{7}{*}{\begin{tabular}[c]{@{}c@{}} 6.96\\6.53\\6.33\\6.01\\ 5.70 \\ \bf 4.17 \\5.31 \end{tabular}}  \\ &           &      &    &    \\ 
 			&           &      &    &    \\
 			&           &      &    &    \\
 			&           &      &    &    \\
 			&           &      &    &    \\
 			&           &      &    &    \\ \hline \hline
 	\end{tabular}	}
 	\label{tab_noise}
 \end{table}
 
For quality comparison, we  {took} the scaling factor  {$s = 8$} for example.  {Fig.~\ref{fig:cave} illustrates the reconstructed images and corresponding error maps of the test image $\emph{Real and fake pepper}$ at 550 nm.} Visually, our method provides the best results in recovering the details of the latent HR HSIs and the corresponding error maps are closer to zero. %{Fig.~\ref{fig:spec_diff} (a)-(c)} show that the spectral differences with the ground truth at three randomly selected points on the image \emph{Real and fake peppers}. It can be observed that the proposed method gives the best approximation of the spectral curves in the latent HR HSIs. 
This demonstrates the efficacy of our method in recovering the spatial information of the latent HR HSIs.

{We  {evaluated} the response curves with the scaling factor $s = 8$ on the CAVE dataset and search their optimal points under MDC.  For each response curve, the hyperparameter $\mu$  {was} sampled on a 50 common logarithmic scale varying form $10^{-6}$ to $1$.  Fig.~\ref{fig:curve} shows the response curves and the points with the optimal parameter values. 
 {Also, we evaluated the RMSE curves by varying the hyperparameter $\mu$ values and the estimated optimal values under MDC. Fig.~\ref{fig:performance} shows the RMSE curves varying with $\mu$ and the estimated optimal points. For better illustration, the horizontal axis is plotted in a common logarithmic scale. As can be observed, when $\mu$ equals the optimal value estimated under MDC, each RMSE curve reaches corresponding minimum point, respectively.}
The optimal parameter values corresponding to different images with different scaling factors are slightly different but unique in our case.}

\subsubsection{Evaluation on Harvard Dataset}
We further  {evaluated} the proposed method on the Harvard dataset. Different from the CAVE dataset, the spatial size of images from the Harvard dataset is larger and slightly moving objects are not correctly aligned in the neighboring bands. As illustrated in Table~\ref{tab_harvard}, with the same experimental setups as discussed above, the numerical results of the proposed method surpass other competing methods in most cases.  {Fig.~\ref{fig:harvard} displays the reconstructed images and corresponding error maps of image \emph{imgf1} at 650 nm band with $s = 8$.} It can be seen that our proposed method reconstructs more details and consequently produces smaller reconstruction error. %As shown in {Fig.~\ref{fig:spec_diff} (d)-(f)}, spectral differences {($s = 16$)} with the ground truth at three randomly selected points on the image \emph{imgf1} clearly demonstrate that the proposed method performs best in estimating the spectral patterns of the latent HR HSIs. 

\subsubsection{ {Evaluation on Chikusei Dataset}}
 {A public hyperspectral dataset with more bands, i.e., the Chikusei dataset, was used to evaluate the performance including the preservation of spectral information of the proposed method. The quantitative metrics of comparing methods with different scaling factors on the Chikusei dataset are reported in Table~\ref{tab_chikusei}. It can be observed that the numerical results of the proposed method surpass other competing methods in most cases. 
Figure~\ref{fig:chikusei} displays the reconstructed images and corresponding error maps of a test image at 800 nm band with $s = 8$. It can be seen that our method reconstructs more details and the corresponding error maps are closer to zero.}

\subsubsection{Evaluation with Optics Blur}
In the above experiments, we  {applied} the uniform blurring matrix $\mathbf{B}$ of size $s \times s$ to each band of the ground truth $\mathbf{X}$ before the down-sampling operator. In the real-world hyperspectral imaging system, however, the optics blur may exist and it can be modeled by a Gaussian blurring kernel. To demonstrate the advantage of the proposed method when optics kernel occurs, we also simulate the LR HSI $\mathbf{Y}$ by first applying $8\times 8$ Gaussian kernel with standard deviation 3 to the latent HR HSI $\mathbf{X}$ before down-sampling in both the vertical and horizontal directions with scaling factor 8.  {The numerical results on the CAVE,  Harvard and Chikusei datasets are reported in Table \ref{tab_gaussian}.} We observe that the proposed method outperforms other considered methods.

\subsection{ {Performance Evaluation under Noisy Conditions}}
 {We conducted some experiments under noise conditions to evaluate the robustness of the proposed method. These experiments were conducted on the CAVE dataset with the scaling factor $s = 8$ and Gaussian blur kernel. We set the SNR to 40 dB to generate noise interruptions, which were added to both two input images. In Table~\ref{tab_noise}, it can be seen that the performance of all the methods declined compared with that in the noise-free condition shown in Table~\ref{tab_gaussian}. However, the proposed method still provided the best quantitative results for the noisy LR HSI and HR conventional image.}
\begin{figure*}[htp]
	\centering
	\includegraphics[scale=1]{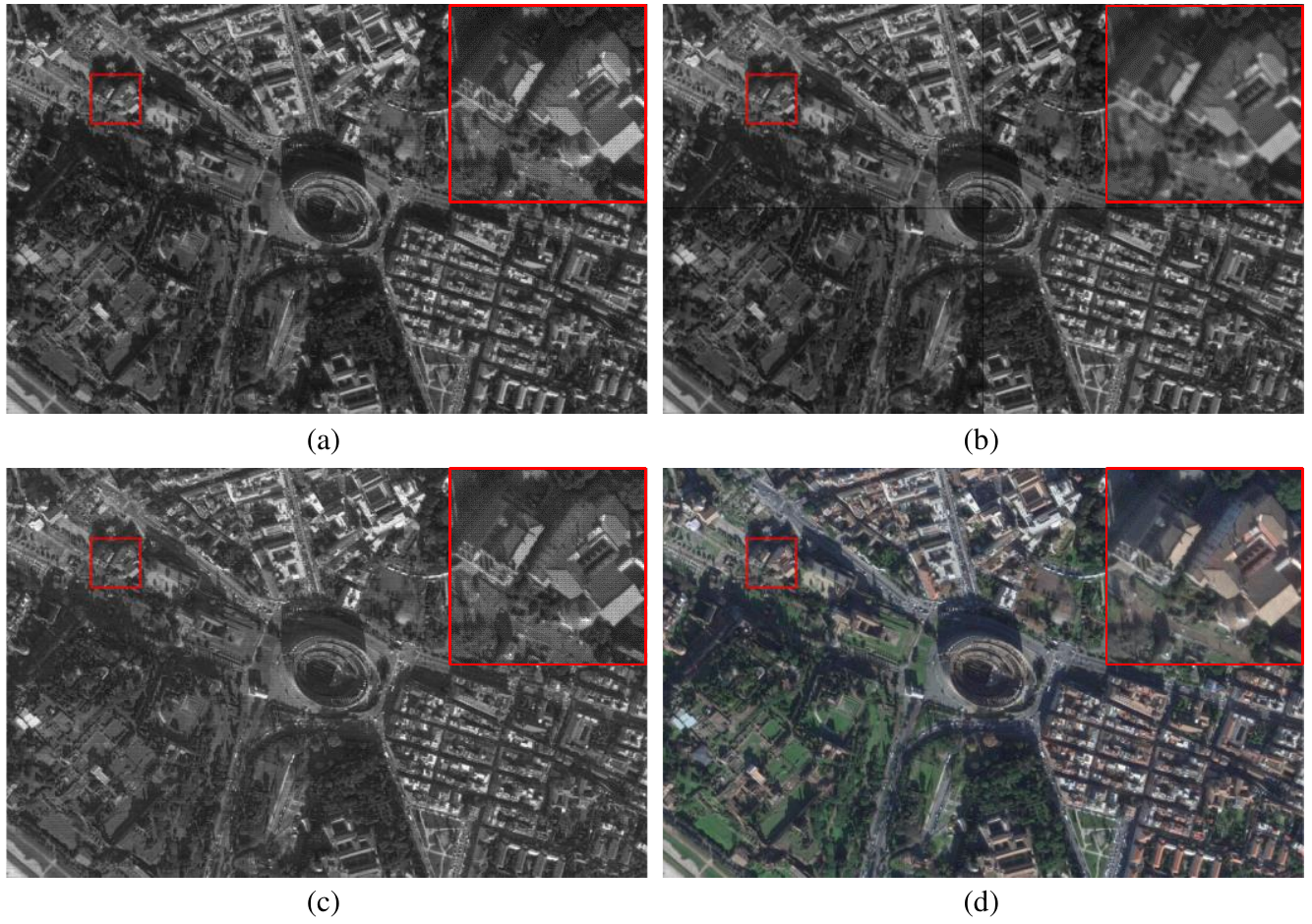}
	\caption{ {Reconstructed images of the real dataset at 605 nm band and and the HR RGB image. Methods: (a) LTTR. (b) UAL. (c) Proposed. (d) HR RGB.}}
	\label{fig:real}
\end{figure*}
\begin{table}[tp]
	\centering
	\caption{ {Average RMSE, PSNR, ERGAS, SAM of some compared methods with the scaling factor $s = 8$ (Gaussian blur kernel) on the CAVE dataset.}} 
	 { \begin{tabular}{c|cccc}
			\hline \hline
			Method              &  RMSE                 &   PSNR         & ERGAS       &      SAM              \\ \hline \hline
			\multirow{4}{*}{\begin{tabular}[c]{@{}c@{}}Bicubic\\ TSFN$\,\mid\ell_2$-norm loss  \\ TSFN w/o skip-con \\ TSFN\end{tabular}} & \multirow{4}{*}{\begin{tabular}[c]{@{}l@{}}  9.63\\ 1.80  \\2.58\\  \textbf{1.52}\end{tabular}} & \multirow{4}{*}{\begin{tabular}[c]{@{}c@{}}  29.42\\ 44.54 \\42.08\\  \textbf{46.01}\end{tabular}} 
			& \multirow{4}{*}{\begin{tabular}[c]{@{}c@{}}  4.220\\ 0.807 \\ 1.092 \\  \textbf{0.674}\end{tabular}} 
			& \multirow{4}{*}{\begin{tabular}[c]{@{}c@{}}  6.67\\ 4.30 \\ 4.87 \\  \textbf{3.81}\end{tabular}} \\
			&             &                &                 &              \\
			&             &                &                 &              \\
			&                   &                   &                   &                   \\ \hline 
			\multirow{2}{*}{\begin{tabular}[c]{@{}c@{}}Proposed w/o TSFN\\ Proposed \end{tabular}} & \multirow{2}{*}{\begin{tabular}[c]{@{}c@{}}  1.69\\ \textbf{1.13}\end{tabular}} & \multirow{2}{*}{\begin{tabular}[c]{@{}c@{}}  45.35\\ \textbf{48.56}\end{tabular}} & \multirow{2}{*}{\begin{tabular}[c]{@{}c@{}}  0.776\\ \textbf{0.512}\end{tabular}} & \multirow{2}{*}{\begin{tabular}[c]{@{}c@{}}  4.10\\ \textbf{3.32}\end{tabular}} \\ 
			&                   &                   &                   &              \\ \hline \hline
	\end{tabular}	}
	\label{tab_ablation}
\end{table}
\subsection{ {Performance Evaluation on Real Data}}
 {We conducted some experiments on real data to evaluate the efficiency of the proposed method. In this dataset, as the ground-truth HR HSI is not available, we generated the training data using the strategy described in Subsection~\ref{sec:dataset}, and compared the proposed method with two most competitive methods, namely, LTTR~\cite{dian2019learning} and UAL~\cite{zhang2020unsupervised}. Due to the limit of GPU memory, we divided input images of UAL into four parts and concatenated the output images back to obtain the final estimated HR HSI. }  
	
 {Figure~\ref{fig:real} shows the reconstructed images of the real dataset at 605nm and the HR RGB image. We can observe that the proposed method recovers more details and gives better visual effect.  This demonstrates the applicability of our method in real-world scenarios. }

\subsection{Ablation Study}
\label{section:ablation_study}
 {We conducted ablation studies on the CAVE dataset. Without loss of generality, we conducted all these experiments with scaling factor 8 and Gaussian blur kernel mentioned above.}

\subsubsection{ {Effect of Different Loss Functions}}
\begin{figure}[tp]
	\centering
	\includegraphics[scale=1]{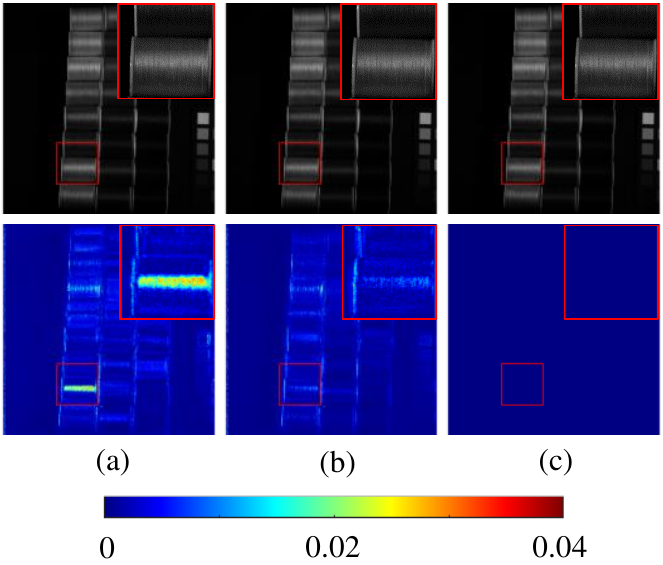}
	\caption{ {Reconstructed images and corresponding error maps of the image \emph{thread spools} from the CAVE dataset with $s = 8$ (Gaussian blur kernel) at 550 nm band. Columns: (a) TSFN($\ell_2$-norm loss) (RMSE: 0.65, PSNR: 51.89). (b) TSFN (RMSE: 0.86, PSNR: 59.41). (c) Ground Truth.}}
	\label{fig:loss}
\end{figure}

 {To compare the performances of $\ell_{1}$-norm and $\ell_{2}$-norm for Eq.~\eqref{eq:loss}, we train the proposed TSFN loss functions with $\ell_{1}$-norm and $\ell_{2}$-norm, respectively. They are termed as ``TSFN" and ``TSFN$\,\mid\ell_2$-norm loss". Table~\ref{tab_ablation} presents the average values of RMSE, PSNR, ERGAS and SAM of these two methods. We can see that $\ell_{1}$-norm shows significant superiority when compared to $\ell_{2}$-norm. The qualitative comparisons are shown in the Figure~\ref{fig:loss}. It visualizes the reconstructed images and corresponding error maps of the test image \emph{thread spools} at 550 nm. Visually, the $\ell_{1}$-norm based loss function provides  better result in recovering the texture details of the latent HR HSI, and produces a smaller reconstruction error.}

\begin{figure}[tp]
	\centering
	\includegraphics[scale=0.2]{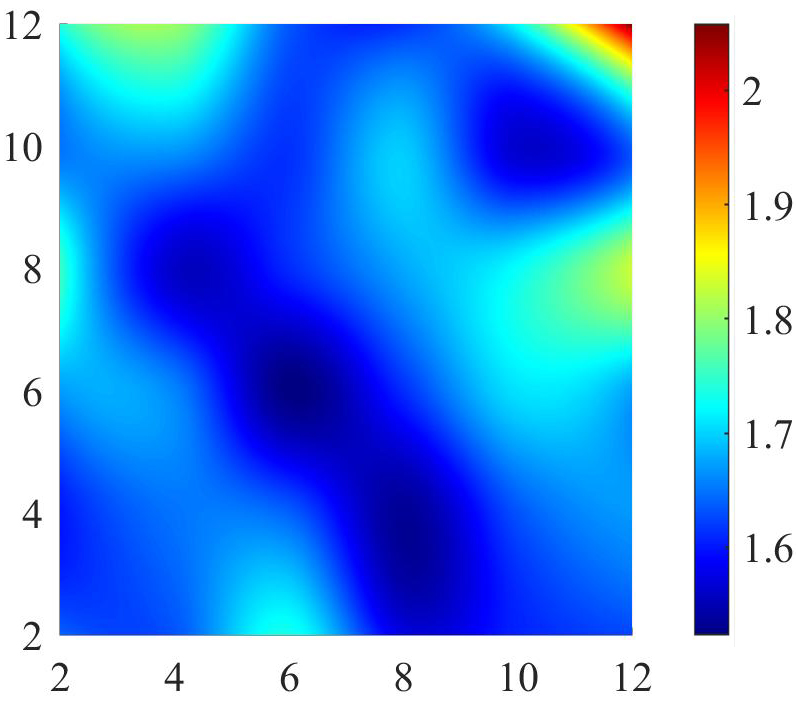}
	\caption{ {2D view of RMSE values with different selections of residual block numbers $P$ and $Q$.}}
	\label{fig:pq}
\end{figure}
\subsubsection{ {Effect of Values of $P$ and $Q$ in TSFN}}
 {The selection of numbers of residual blocks in two streams ($P$ and $Q$) are crucial in our network architecture design. The average RMSE values on test images were used to evaluate the performance. Figure~\ref{fig:pq} shows the two demensional (2D) view of RMSE values with different settings of $P$ and $Q$. Obviously, small values of $P$ and $Q$ will cause a poor inference ability of the network, yielding worse performance. On the other hand, the increase of numbers of residual blocks for feature extraction does not promise better results. It can be observed that the best performance occurred when the values of $P, Q$ are both set around 6.}
\subsubsection{ {Effect of the Skip Connection in TSFN}}
 {The skip connection operator is the key component of our proposed TSFN. To show the effect of this operator, we utilized a variant of TSFN that removes the skip connection operator, which is termed as ``TSFN w/o skip-con". As reported in Table~\ref{tab_ablation}, removing the skip connection operator deteriorates the performance.}
\subsubsection{ {Effect of TSFN in the Proposed Framework}}
As stated in Section~\ref{section:proposed_method},  we propose to learn the priors of the latent HR HSI via TSFN rather than using handcrafted regularizers. Here, we illustrate the effectiveness of TSFN, as well as its important role in our HSI super-resolution framework. First, we  {compared} the up-sampled images $\mathbf{Y}_\text{up}$ with the bicubic interpretation and $\widetilde{\mathbf{X}}$, the outputs of our proposed TSFN. In addition, we  {evaluated} the results of the proposed HSI super-resolution framework without TSFN  {(termed as ``Proposed w/o TSFN'')}. More specifically, we  {used} the up-sampled images $\mathbf{Y}_\text{up}$ as the approximation of the latent HR HSI, and compared with the result of our complete framework with TSFN. 

%We conduct all these experiments on both the CAVE dataset with Gaussian blur kernel mentioned above.

As shown in Table \ref{tab_ablation}, the results of TSFN significantly outperformed those of the bicubic interpretation. This illustrates the effectiveness of the priors representation of our method learned via TSFN. Moreover, the performance of the proposed framework with TSFN exceeds that of the framework without TSFN, thus demonstrating TSFN is an indispensable part of our proposed HSI super-resolution method.

\section{Conclusion}
In this paper, we presented a deep priors-based HSI super-resolution method. Instead of using handcrafted image prior structures, we utilized the spatial-spectral priors learned by the proposed TSFN. The output of TSFN was leveraged to regularize the ill-posed problem. In addition, the regularization parameter was automatically estimated by {adopting MDC on the response curve of the associated problem with a golden-section search approach.} Experimental results on  {both simulated and real data} demonstrated that our method can effectively handle scenarios with various scaling factor and blurring kernel setups. 

In future works, the proposed HSI super-resolution method will be further extended in two directions. On the one hand,  {modeling the data fidelity term by estimating the blurring matrix and the spectral response function using deep learning techniques} will be of interest. {On the other hand, how to accelerate hyperparameter tuning will be investigated.} These attempts will further enhance the applicability of our proposed method in real-world scenarios.

%\newpage

\bibliographystyle{IEEEtran}
\bibliography{mybib}
\end{document}